\documentclass{article}	
\usepackage{hyperref}
\usepackage{url}
\usepackage[utf8]{inputenc} 
\usepackage{geometry} 
\usepackage{amsmath, amssymb, graphicx,times,bbm,algorithm,algorithmic,color,subfigure}
\newcommand{\EE}{\mathbb{E}}

\newcommand{\PP}{\mathbb{P}}
\newcommand{\RR}{\mathbb{R}}
\newcommand{\indic}{{\bf 1}}

\newcommand{\bp}{\noindent{\textbf{Proof.}}\ }
\newcommand{\ep}{\hfill $\Box$}

\newcommand{\als}[1]{ \begin{align*} #1  \end{align*}}
\newcommand{\eqs}[1]{ \begin{equation*} #1  \end{equation*}}
\newcommand{\sk}{\nonumber\\}

\newcommand{\el}{\end{flushleft}}
\newcommand{\bl}{\begin{flushleft}}

\newtheorem{proposition}{Proposition}
\newtheorem{theorem}{Theorem}
\newtheorem{lemma}{Lemma}
\newtheorem{corollary}{Corollary}
\newtheorem{remark}{Remark}

\newtheorem{assumption}{Assumption}

\newcommand{\ei}{ \end{itemize} }
\newcommand{\ben}{ \vspace{-0.15cm} \begin{enumerate} \setlength\itemsep{-0.1cm} }
\newcommand{\tto}[1]{\underset{#1}{\to}}
\newcommand{\bi}{ \vspace{-0.15cm} \begin{itemize} } %\setlength\itemsep{-0.1cm} }
\newcommand{\een}{ \end{enumerate} }

\newcommand{\figsize}{1 \columnwidth}
\newcommand{\figsizeA}{0.49\columnwidth}
\newcommand{\lbracket}{[\![}
\newcommand{\rbracket}{]\!]}

\begin{document}
\title{Multipath Streaming: Fundamental Limits And Efficient Algorithms}
\author{Richard Combes$^{\dag}$,  Habib Sidi$^{\times}$ and Salah-Eddine Elayoubi$^{\ddag}$ 
\thanks{$\dag$: Centrale-Supelec and L2S, Telecommunication Department, Gif-sur-Yvette (France), {richard.combes@centralesupelec.fr}}
\thanks{$\times$: INRIA, Paris (France), {habisd1@gmail.com}}
\thanks{$\ddag$: Orange Labs \& IRT SystemX, Paris (France), {salaheddine.elayoubi@orange.com}}
}
\renewcommand\footnotemark{}
\renewcommand\footnoterule{}
\maketitle

\begin{abstract}
We investigate streaming over multiple links. A file is split into small units called chunks that may  be requested on the various links according to some policy, and received after some random delay. After a start-up time called pre-buffering time, received chunks are played at a fixed speed. There is starvation if the chunk to be played has not yet arrived. We provide lower bounds (fundamental limits) on the starvation probability of any policy.  We further propose simple, order-optimal policies that require no feedback. For general delay distributions, we provide tractable upper bounds for the starvation probability of the proposed policies, allowing to select the pre-buffering time appropriately. We specialize our results to: (i) links that employ CSMA or opportunistic scheduling at the packet level, (ii) links shared with a primary user (iii) links that use fair rate sharing at the flow level. We consider a generic model so that our results give insight into the design and performance of media streaming over (a) wired networks with several paths between the source and destination, (b) wireless networks featuring spectrum aggregation and (c) multi-homed wireless networks. 
\end{abstract}
%{\bf Keywords}: Streaming; Performance Evaluation; Video Streaming; Wireless Networks; Wired Networks; Mulithoming; Spectrum Aggregation.
\section{Introduction}

We consider the problem of streaming a file divided into small units called ``chunks'' over multiple links.  Each chunk may be requested on any link. Requested chunks arrive in the requested order, separated by random delays. Randomness is due to propagation delays, change in the link quality and congestion since the streaming flow shares the links with competing flows. Received chunks are stored in a buffer. After a fixed start-up time called pre-buffering time, the received chunks are played at a fixed speed. If a chunk does not arrive before the time it is supposed to be played there is starvation. For video streaming, starvation causes interruptions in the play-out and is the main cause for poor Quality-of-Service (QoS). The goal is to design chunk request policies that minimize the starvation probability.

For clarity we use the generic term ``links'' to denote entities on which packets may be requested. Links may represent for instance: (a) several disjoint paths between a source and a destination in a wired network (b) several disjoint frequency bands in a wireless system featuring spectrum aggregation, (c) several wireless interfaces which employ different protocols, for instance cellular data, say LTE (Long Term Evolution) and local area, say 802.11 (WiFi). Case (c) is often referred to as ``multihoming''. While links may have different data rates as well as different traffic management rules, a good protocol should ensure that, from the point of view of the streaming flow, they appear as a single aggregated entity whose data rate is the sum of the links data rates. The goal is hence to perform link aggregation efficiently so as to minimize the starvation probability.

The problem is motivated by several features of current networks. First, in current mobile networks, video-on-demand services consume the majority of the available bandwidth, and good traffic management rules have a dramatic impact on the network performance. Further, due to the massive deployment of both cellular data networks and local area networks, most users in dense urban zones have access to several wireless links. If feasible, spreading the load evenly on those links is highly desirable. Finally current streaming protocols such as the DASH standard~\cite{Stockhammer11} operate by splitting video files in chunks and requesting them as independent HTTP requests. There are two advantages to this approach: (i) this is a stateless protocol, the video server does not keep track of the state of the user playing the video, and processes each chunk request independently (ii) requesting each chunk through HTTP avoids issues with middle-boxes (e.g. firewalls and NATs) which may block certain ports for security.

There are two challenges. First the number of request policies is exponential in the number of chunks, so that exhaustive search is not feasible. Furthermore, one would like to minimize the amount of feedback required by policies to ease implementation. The second challenge is to  calculate the starvation probability of a given policy by tractable formulas. This enables to select the pre-buffering time to ensure that the starvation probability is smaller than some threshold. For elastic traffic one is concerned with the maximum achievable long-term throughput, which depends mostly on the expected delay but not on the full delay distribution. On the other hand for streaming traffic, starvation is highly dependent on the variability of the delays (e.g. the jitter), and accurate predictions of starvation probability must depend on the full delay distribution. Analytical formulas are useful even when one can simulate the system since, when the pre-buffering time is large, starvation tends to be a rare, ``large-deviations'' type of event. So one needs long simulations to get accurate estimates.

{\bf Our contribution}: We first derive lower bounds on the starvation probability of any policy. Those bounds enable us to derive the scaling of the pre-buffering time required to reach a target starvation probability. We further propose simple, implementable policies (without feedback) to determine how chunks may be requested on the various links. We derive upper bounds and/or approximations for the starvation probability of these policies. Those upper bounds are tractable, including both cases where delays are independent and correlated. We deduce that the proposed  policies are efficient since they approach the fundamental limits derived earlier. Finally we specialize our results to several situations of practical interest: (i) links that employ opportunistic scheduling or random access at the packet level (ii) links where the streaming flow is the secondary user of a cognitive radio network, and is allowed to use the link only when the primary user is inactive (iii) links where the streaming flow shares the available bandwidth with short-lived flows using (for instance) fair rate sharing.

The remainder of this article is organized as follows. In section~\ref{sec:related_work} we highlight the relationship between our results and previous work. In section~\ref{sec:model} we state our assumptions and notation.  We provide performance limits and state the proposed schemes in section~\ref{sec:limit_algo}. In section~\ref{sec:performance} we calculate the performance of the proposed schemes under general assumptions. We specialize our results to several popular link models in  section~\ref{sec:case_studies}. In section~\ref{sec:numerical} we illustrate our results through numerical experiments. Section~\ref{sec:conclusion} concludes the article.

\section{Related work}\label{sec:related_work}
\subsection{Performance evaluation and control for streaming}

Streaming over wireless networks has become popular, and a lot of attention has been dedicated to its performance evaluation, to understand the relationship between pre-buffering and starvation. \cite{TMM10:Luan} considers a G/G/1 queue where the first two moments of the arrival and service distributions are known. The discrete buffer size is approximated by a Wiener process. \cite{TMM08:Liang} considers an ON-OFF  wireless channel and performs a probabilistic analysis based on the knowledge of the playback and arrival curves. An M/M/1 queue is used in \cite{Infocom12:Xu} to derive explicit formulas for the number of starvations, based on the Ballot theorem~\cite{Takacs}. \cite{YuedongInfocom13} considers the combination of fading and flow level dynamics, and calculates the starvation probability by solving a set of ordinary differential equations constructed on top of the Markov chain describing the flow level dynamics. 

All these works considered a single wireless link in a classical TCP/IP architecture. Another set of works studied the impact of multihoming on streaming QoS. For instance, \cite{habib} studies multihomed streaming in a residential context using a DSL and a cable connection. Experiments show that connection splitting/migration in case of congestion provides a significant QoS improvement. For lossy wireless channels, \cite{JurcaRateAllocation} considers the rate allocation problem as an optimization problem to minimize a distortion metric for the video taking into account the loss rate of each of the network paths. \cite{JurcaFEC} extends this work to joint scheduling and Forward Error Correction (FEC), while \cite{multihome_WiFi_UMTS} considers the specific case of WiFi/UMTS multihoming in the design of the FEC scheme. The same objective of reducing the video distortion is used in \cite{multihomed_TON} to design a multipath scheduling policy. These works aim at reducing video distortion, while we are interested in HTTP-like video streaming with pre-buffering where video chunks are downloaded without distortion and the QoS metric is the starvation probability.

Another set of works on multihomed streaming is related to content centric networking where the rate allocation between paths is driven by interest packets generated by the users. Authors in~\cite{salsano_streaming} consider a group of neighboring users using their mobile devices to view the same video, and exploit the fact that users can exchange video chunks using direct connections instead of downloading all the chunks using the cellular link. Others consider {\em real time} services with stringent delay constraints for instance voice \cite{Jacobson_voice} and live TV \cite{Xu_TV}, which is out of the scope of the present work. 

To the best of our knowledge, the present work is the first to derive optimal allocation strategies for streaming over multi-homed access with the starvation probability as the main performance indicator and provide strong theoretical performance guarantees given in the form of tractable, analytical formulas.

Two contributions on multipath streaming which are complementary to ours are \cite{Singh13} and \cite{Xing14}. In \cite{Singh13}, the authors propose a protocol for multipath streaming and evaluate its performance experimentally but do not conduct a theoretical analysis. The authors of \cite{Xing14} formulate the problem as a Markov Decision Process (MDP). Since computing the optimal policy of this MDP is prohibitively expensive they propose a heuristic policy search algorithm for which no theoretical guarantees are available and its performance is evaluated numerically.

\subsection{Receiver driven transport protocols}
While IP transport is under destination control, several receiver-driven transport protocols have been proposed (see Gupta et al. in \cite{gupta1998webtp}). Receiver driven protocols have recently attracted a lot of attention with the development of Information-Centric Netwroking (ICN), where download rates are determined by the rate at which interests are sent into the network by the end-system. Arianfar et al. \cite{arianfar2010contug} discuss the challenges faced when designing transport protocols for content centric networks and present a receiver-driven transport protocol, ConTug, that takes into account the presence of multiple sources. Carofiglio et al. propose another receiver-based protocol called ICP (Interest Control Protocol) using AIMD window-based congestion control \cite{Carofiglio2012a,Carofiglio2013b}. \cite{Saino2013} proposes CCTCP (Content Centric TCP) that considers the use of multiple sources in the design of the protocol parameters such as the timeout timer.

In our work, we are neither interested in the design of such transport protocols, nor in the evaluation of their performance for streaming traffic. Our objective is to design optimal policies for splitting interests over the different interfaces. We nevertheless assume that mechanisms that adapt the interest sending rates to the state of each multihoming path are implemented, ensuring that the available capacity of each link is fully used. Our policies can thus be implemented at the upper layers over classical TCP/IP based protocols, or in the form of a receiver-driven interest control protocol. 

\subsection{Spectrum Aggregation}
Spectrum aggregation is driven by two different but complementary objectives: achieving higher user data rates and exploiting all the available spectrum to offload traffic from the mobile network. Typical solutions to attain the former objective are the so-called carrier aggregation features in 3GPP, starting with the dual cell feature in 3G systems (Release 8). The dual cell feature aggregates two 3G carriers of 5 MHz and doubles the achievable data rate~\cite{Morais09}. However, the dual cell feature has been shown (\cite{Bonald09}) to achieve significant capacity gains for streaming services due to the trunking effect, especially since the data rate in Release 7 is not sufficient to meet the requirements of TV streaming . It is noted that this gain is reduced when MIMO is combined with the dual cell feature in Release 9. The term ``dual-cell'' has been changed to ``carrier aggregation'' in recent standards as more than two carriers are aggregated starting from Release 10. 3GPP then introduced a more advanced carrier aggregation mechanism, where the unused spectrum in the 1.4 GHz band (originally allocated for TDD mode but not used in Europe) is aggregated with the FDD LTE spectrum via multihoming \cite{3GPP}.

Another set of spectrum aggregation technologies aims at increasing the system capacity and offloading the cellular traffic to other systems, while offering a better QoS to users. This started with Cognitive Radio (CR) which suggests to reuse unused spectrum allocated to other services in a secondary, non-priority, way \cite{Mitola99}. The first commercial services related to CR reuse white spaces left by TV systems \cite{Nekovee09}. However, as spectrum availability cannot be guaranteed to secondary users, multihoming is advocated: users simultaneously connect to both licensed and non-licensed spectrum. 3GPP systems adopt multihoming to aggregate 4G spectrum with WiFi spectrum, in the so-called dual connectivity mode \cite{Zakrzewska13}, exploiting the presence of both WiFi and 4G radio frequency modules in almost all user devices. The 3GPP LWIP Release 13 standard includes aggregation of WiFi and LTE (called ``Wifi Boost'') at the IP layer, and \cite{Ling15,Perez16} show that large performance gains result from this aggregation. 

As opposed to classical carrier aggregation where resource allocation is fully centralized,  systems based on CR and dual connectivity can be user-centric: users sense the medium and decide which packets to send to each carrier, possibly assisted by the network \cite{Dandachi16}. Decisions are taken by users based on local information (sensing) and contextual information (link quality estimates) provided by the network. Different carriers are usually heterogeneous in terms of radio channel quality (independent fast fading), coverage (path loss and slow fading) and load (number of connected users). The decisions taken by users have a large impact on QoS and an efficient protocol is needed.

\section{The model}\label{sec:model}

\subsection{Basic Description}

We consider a file divided in $N$ chunks of unit size, indexed by $n \in \lbracket 1,N \rbracket$. There are $K \geq 1$ links, on which any chunk can be requested. When a requested chunk is received on a link, it is placed in a buffer. After a pre-buffering time denoted by $B > 0$ the file is read at unit speed. Namely, at time $n + B$, if chunk $n$ is present in the buffer then it is read, and otherwise starvation occurs.

We define the chunk request policy which is represented by a vector $\pi \in \lbracket 1,K \rbracket^N$, where $\pi_n = k$ if chunk $n$ is requested on link $k$. We assume that if two chunks $n < n'$ are requested on the same link then $n$ is requested before $n'$. It is clear that this always ensures lower starvation. We define $d_k(n) = \sum_{n'=1}^{n} \indic\{ \pi_{n'} = k \}$, the number of chunks comprised between $1$ and $n$ requested on link $k$. We further define $X_k(\ell)$ the delay of the $\ell$-th chunk requested on link $k$. Namely, if $\pi_n = k$, chunk $n$ arrives at time $\sum_{\ell = 1}^{d_k(n)} X_k(\ell)$. The starvation probability $P^N$ is the probability that there exists a chunk that arrives after the time it was supposed to be played, so that:
\eqs{
P^N = \PP\left[ \exists k, n : \pi_n = k \text{ and } \sum_{\ell=1}^{d_k(n)} X_{k}(\ell) > n + B\right].
}
It is noted that in general $\pi$ can be sample path dependent, so that $\pi$ depends on the delays $( X_k(\ell) )_{\ell,k}$. We will consider two types of policies.
\bi
\item \underline{Static policies}: $\pi$ does not depend on $( X_k(\ell) )_{\ell,k}$
\item \underline{Oracle policies}: $\pi$ is an arbitrary function of $( X_k(\ell) )_{\ell,k}$
\ei
Static policies may depend on the delay statistics but not on their realization, so that they require no feedback and are attractive in terms of implementation. Oracle policies are policies that ``know everything'', including the delays of chunks not yet received. Oracle policies are not implementable and mainly serve here as a performance upper bound. 
\subsection{Assumptions}
We introduce several sets of assumptions on the delays.
\begin{assumption}[I.I.D. delays] \label{as:iid}
For all $k$, $(X_k(\ell))_{\ell \geq 0}$ is an i.i.d sequence with expectation $\mu_k$, variance $\sigma_k^2$ and cumulant generating function $G_k(a) = \log( \EE[e^{a X_k(\ell)}] )$. We assume that $G_k(a) < +\infty$ on an open neighbourhood of $0$.
\end{assumption}
\begin{assumption}[Markovian delays] \label{as:markov}
	For all $k$ there exists a continuous time, stationary ergodic Markov chain on a discrete space ${\cal S}$ denoted by $( S_k(t) )_{t \in \RR}$ with stationary distribution $m_k$ and transition rate matrix $Q^k = (q^k(i,j))_{i,j \in {\cal S}}$. There exists a function $r: {\cal S} \to \RR^+$ such that for all $\ell,k$:
\als{
\tau_k(\ell) &= \min \left\{  t \geq 0: \int_{0}^{t} r(S_k(u)) du \geq \ell \right\}, \sk
 X_k(\ell) &= \tau_k(\ell) - \tau_k(\ell-1).
}
We define $\mu_k = \EE[X_k(\ell)]$.
\end{assumption}
The Markovian case can be understood as follows: at time $u$, link $k$ is in state $S_k(u)$ and the streaming flow has an instantaneous data rate of $r(S_k(u))$. So  $\tau_k(\ell)$ is the duration required to receive $\ell$ chunks of unit size. Hence $X_k(\ell) = \tau_k(\ell) - \tau_k(\ell-1)$. The state is typically the number and/or state of the other flows sharing link $k$ with the streaming flow. For both sets of assumptions we define $r_k = 1/\mu_k$, the average data rate of link $k$, and $R = \sum_{k=1}^K r_k$ the sum of data rates. Based on the value of $R$ we distinguish three regimes: underload if $R > 1$, critical if $R = 1$ and overload if $R < 1$. We define the frequency vector $f = (f_1,...,f_K)$, with $f_k = r_k/R$. In order to balance the loads of the available links, the number of chunks requested on link $k$ should be close to $N f_k$, hence the name "frequency" for $f_k$. We shall see that the proposed algorithms precisely accomplish that goal. We denote by $\Psi(x) = \frac{1}{\sqrt{2 \pi}} \int_{x}^{+\infty} e^{-\frac{z^2}{2}} dz$ the complementary c.d.f of the standard Normal distribution. For clarity, we denote by $P^N(\pi,B)$ the starvation probability for $N$ chunks, prebuffering time $B$ and policy $\pi$. Finally we introduce an assumption on the independence of delays across links. 
\begin{assumption}[Independent Links] \label{as:links}
$(X_k(\ell))_{\ell \geq 0}$ is independent from $(X_{k'}(\ell))_{\ell \geq 0}$ for all $k \ne k'$.
\end{assumption}
\subsection{Applicability of the model and rate adaptation}
Two remarks about the model are in order. First it should be noted that the Markovian delay assumption includes \emph{any} delay distribution and \emph{any} form of correlation between the successive delays observed on a given link. Indeed, there are no restrictions on the size of the state space $\cal S$ nor on the transition rates $Q^k$. For instance the process $(S_k(t))_k$ can follow any of the classical queues (as considered in prior work) say M/M/1, G/G/1 etc. 

Second, throughout the article we assume that the video data rate is fixed and it is taken to be unity in our notation simply to ease presentation. Indeed, in practice, some form of rate adaptation will be implemented: before the video is played, the data rates $r_1,...,r_K$ will be estimated, and the data rate of the video will be chosen approximately equal to $\sum_{k=1}^K r_k$. This guarantees that the video quality is as high as possible while starvation probability can be made small by choosing the pre-buffering time appropriately and using an efficient policy to request chunks.

Performing rate adaptation is relatively simple since it involves only first order statistics, namely the first moments of the delays $\mu_k$, since $r_k = 1/\mu_k$. On the other hand selecting the pre-buffering time to avoid starvation seems more challenging since it involves the complete distribution of delays including higher moments (e.g. the jitter), as starvation is mostly a "tail event" due to large, rare fluctuations of the delays.

\section{Performance limits and algorithms}\label{sec:limit_algo}
\subsection{Fundamental Limits}
 Our first result presented in theorem~\ref{th:iid_lower_bound} is a lower bound on the starvation probability that holds for all oracle policies (hence for static policies as well). The intuition behind statement (i) is simple: if starvation does not occur, then, for all $n$ the total number of chunks received on all links at time $n + B$ must be greater or equal to $n$, otherwise there exists a chunk $n'$ which has not been received at time $n' + B \leq n + B$. The lower bound is not explicit but may be computed easily by simulation. Statement (ii) shows that for large files ($N \to \infty$), if $R \leq 1$, then the pre-buffering time must be of the order $(R^{-1} - 1)N + {\cal O}(\sqrt{N})$, otherwise the starvation probability tends to $1$.  For large files, we obtain an explicit lower bound using the central limit theorem. This lower bound is an increasing function of the delay variances $\sigma_k^2$, and tends to $1$ when we let $\min_k \sigma_k^2 \to \infty$. As a consequence, in the critical case, if the delays are heavy tailed, one cannot ensure a starvation probability $< 1$ with ${\cal O}(\sqrt{N})$ pre-buffering time. Also there are sharp transitions between the regimes: from $B = {\cal O}(1)$ (underload) to $B = {\cal O}(\sqrt{N})$ (critical) to $B = {\cal O}(N)$ (overload).  We will subsequently show that these orders are tight. Also note that, intuitively, the law of iterated logarithm would suggest the pre-buffering time to be at least ${\cal O}(\sqrt{N \log\log(N)})$ in the critical regime and we show that this intuition is flawed.

\begin{theorem}\label{th:iid_lower_bound}
	The following holds for all oracle policies $\pi$. 
	
	(i) For all $B \geq 0$ and $N \geq 0$ we have:
	\als{
	P^N(\pi,B) &\geq \PP \left[ \exists n \in \lbracket 1,N \rbracket : \sum_{k=1}^K D_k(n) < n  \right] \sk
	D_k(n) &= \max\{ d \geq 0: \sum_{\ell=1}^{d} X_k(\ell) \leq B + n\}.	
	}
	(ii) Let assumptions~\ref{as:iid} and \ref{as:links} hold. If $R \leq 1$, for all $b \geq 0$:
	\als{
		\lim \inf_{N \to \infty} P^N(\pi, (R^{-1} - 1)N + b\sqrt{N} ) &\geq \prod_{k=1}^{K} \Psi\left( \frac{b}{ \sigma_k \sqrt{ f_k }} \right).
	}
\end{theorem}

\bp
(i) Consider a fixed oracle policy $\pi$. Define the event 
\eqs{{\cal A} = \left\{ \exists n \in \lbracket 1,N \rbracket : \sum_{k=1}^K D_k(n) < n  \right\}.
} 
Consider $n$ fixed and assume that $\sum_{k=1}^K D_k(n) > n$. Define $N^{\pi}_k(n)$ the number of chunks received on link $k$ before time $B + n$, and $N^{\pi}(n) = \sum_{k=1}^K N_k^{\pi}(n)$ the total number of chunks received before time $B + n$. By definition $N^{\pi}_k(n) \leq D_k(n)$ for all $k$, so that $N^{\pi}(n) \leq \sum_{k=1}^K D_k(n) < n$. Hence, at time $n + B$, there exists a chunk $n' \leq n$ which has not been received. We deduce that if ${\cal A}$ occurs, starvation occurs which proves the first claim.

(ii) Define $B^N =  (R^{-1} - 1)N + b\sqrt{N}$. Define the following events:
\eqs{
{\cal B}^N_k = \{ D_k(N) < N f_k \} \;\;,\;\; {\cal B}^N = \cap_{k=1}^K {\cal B}_k.
} If ${\cal B}^N$ occurs, $\sum_{k=1}^K D_k(N) < N \sum_{k=1}^K f_k = N$  so that starvation occurs.  Event ${\cal B}^N_k$ occurs if and only if \eqs{
\sum_{\ell=1}^{N f_k} X_k(\ell) > N + B^N,
} 
so that, replacing $B^N$ by its expression:
\eqs{
	\frac{1}{\sqrt{N f_k}} \sum_{\ell=1}^{N f_k} (X_k(\ell) - \mu_k) > \frac{N + B^N - N f_k \mu_k}{\sqrt{N f_k}} = \frac{b}{\sqrt{f_k}}. 
}
By the central limit theorem:
\eqs{
\frac{1}{\sqrt{N f_k}} \sum_{\ell=1}^{N f_k} (X_k(\ell) - \mu_k) \tto{N \to \infty} {\cal N}(0 , \sigma_k^2),
}
in distribution so that:
\eqs{
	\PP[{\cal B}_k^N] \tto{N \to \infty} \Psi\left( \frac{b}{ \sigma_k \sqrt{ f_k }} \right).
}
Events ${\cal B}_1^N,...,{\cal B}_K^N$ are independent so that
\eqs{
P^N(\pi, B^N) \geq \prod_{k=1}^K \PP[{\cal B}^N_k] \tto{N \to \infty} \prod_{k=1}^K \Psi\left( \frac{b}{ \sigma_k \sqrt{ f_k }} \right).
}
which gives the announced result.
\ep

\subsection{Efficient Algorithms}
Intuitively, to obtain an efficient policy, chunks should be requested on link $k$ at frequency $f_k$, so that the number of requested chunks $d_k(n)$ is close to $n f_k$. We define a class of static policies called upper balanced policies. We say that $\pi$ is $f$-upper balanced if for all $k, n$ we have:
\eqs{
	d_k^\pi(n) \leq (n + K - 1) f_k .
}
If frequencies are not rational numbers it is not obvious how to build such a policy. We show a simple recursive way to construct upper balanced policies with arbitrary frequencies.
\begin{proposition}\label{prop:up}
	Consider $\pi$ such that for all $n \geq 0$: 
\eqs{
	\pi_{n} \in \arg \min_{k} \frac{d_k^\pi(n-1) + 1}{f_k}
}
with ties broken arbitrarily. Then $\pi$ is $f$-upper balanced.
\end{proposition}
\bp
	We drop ${.}^\pi$ for convenience. Define $a(n) = \frac{d_{\pi_{n}}(n-1) + 1}{f_{\pi_{n}}}$. Since $\pi_{n} \in \arg \min_{k} \frac{d_k^\pi(n-1) + 1}{f_k}$, we have  $a(n) \leq \frac{d_{k}(n-1) + 1}{f_{k}}$ for all $k$. Hence $f_k a(n) \leq d_{k}(n-1) + 1 $ for all $k$, and summing over $k$ we get: 
$a(n) = a(n) \sum_{k} f_k \leq \sum_{k} (d_{k}(n-1) + 1)  = n - 1 + K$. 
	
 Let us now prove that $d_k(n) \leq  (n + K - 1) f_k$ for all $n$. We proceed by induction. The inequality is true for $n = 0$ since $d_k(0) = 0 \leq (K - 1)f_k $.  Consider $n$ such that $d_k(n-1) \leq (n - 1 + K - 1)f_k$ for all $k$. If $k \neq \pi_n$ then $d_k(n) = d_k(n-1) \leq (n - 1 + K - 1)f_k \leq (n + K - 1)f_k$. On the other hand, if $k = \pi_n$ we have $d_k(n) =  d_k(n-1) + 1 = f_k a(n) \leq ( n + K - 1) f_k$. Therefore $d_k(n) \leq  (n + K - 1) f_k$ for all $n$ and all $k$ as announced.
\ep

Upper balanced sequences share a close resemblance with balanced sequences used in optimal routing problems such as \cite{Hajek85}.
A key difference is that it is usually difficult to compute balanced sequences when either $K \geq 3$ or frequencies are not rational. When frequencies are not rational, one cannot construct upper balanced sequences that are periodic, so that a (simple) recursive formula to calculate upper balanced sequences seems good enough. It is also noted that, contrary to problems of maximal throughput scheduling for elastic traffic, Bernoulli routing is not satisfactory. Namely using randomized sequences $\pi$ where $(\pi_n)_n$ is an i.i.d sequence on $\{1,...,K\}$ with distribution $(f_1,...,f_K)$ is no good since this would give $d_k^\pi(n) - n f_k = {\cal O}(\sqrt{n})$ (central limit theorem).

The policy defined in proposition~\ref{prop:up} can be implemented efficiently in practice as follows. As said previously, rate adaptation is performed so that $r_1,...,r_K$ and $R$ are estimated before chunks are requested. This allows to estimate frequencies $f_1,...,f_K$, and in turn calculate $\pi_n$ (the link on which chunk $n$ is requested) for all $n$ using proposition~\ref{prop:up}. This requires ${\cal O}(N)$ operations and ${\cal O}(K)$ memory.   An advantage of this policy is that it is completely static, hence it does not require to monitor the state of the various links once the streaming flow has started.
 
For very long files, the data rates of the various links may change on a slow time scale (say a few minutes), and to deal with this one may simply perform rate adaptation periodically and recompute $\pi$ based on proposition~\ref{prop:up}.

\section{Performance evaluation}\label{sec:performance} 
\subsection{Performance for i.i.d. delays}
We consider i.i.d. delays. Our second result is theorem~\ref{th:iid_upper}, which gives upper bounds for the starvation probability of upper balanced policies, and shows that, in the 3 regimes of interest they are order optimal. Namely the pre-buffering times have the same scalings as the lower bound of theorem~\ref{th:iid_lower_bound}. More precisely, for statement (iii), we consider the regime where $R \le 1$. From our previous result the pre-buffering time must be greater than $(R^{-1} - 1) N  + {\cal O}(\sqrt{N})$, otherwise $P \to 1$ when $N \to \infty$. So we consider pre-buffering time $(R^{-1} - 1) N  + b + K-1$ and study the starvation probability as a function of $b$. Indeed, $P \not\to 1$ if and only if $b$ scales as ${\cal O} (\sqrt{N})$ when $N \to \infty$.

The proof is based on Doob's maximal inequality and before stating theorem~\ref{th:iid_upper} we define the exponent $a_k^\star$ which appears in our upper bounds. We define $F_k(a) = G_k(a) - a/f_k$. If $X_k(\ell) < 1/f_k$ a.s. then define $a_k^\star = +\infty$, and otherwise define $a_k^\star = \max \{ a \geq 0: F_k(a) = 0 \}$. Given $G_k(.)$, $a_k^\star$ can always be calculated numerically using a zero-finding procedure such as bisection or Newton's method. We further give explicit formulas for calculating $a_k^\star$ in several cases of interest: exponential delays and sub-Gaussian delays, see Proposition~\ref{prop:subgaussian}. We recall that $X_k(\ell)$ is sub-Gaussian if there exists $v^2_k \geq 0$ such that for all $a \geq 0$: $G_k(a) \leq a\mu_k + a^2 v_k^2/2$. It is noted that if delays are  Gaussian or bounded, then they are sub-Gaussian, see Remark~\ref{rem:subgaussian}.
\begin{proposition}\label{prop:subgaussian}
Consider $R > 1$. 

(i) If $X_k(\ell) \sim$ Exp$(r_k)$, then $a^\star_k = r_k(1 + W( -R e^{-R})/R)$ with $W$ the Lambert function.

(ii) If $X_k(\ell)$ is $v^2_k$-sub-Gaussian then $a^\star_k \geq 2 \mu_k (R-1) / v^2_k$.
\end{proposition}
\bp
	(i) If $X_k(\ell) \sim$ Exp$(r_k)$, then $G_k(a) = r_k/(r_k - a)$, $a \in [0,r_k)$. So $a^\star_k$ is a solution to the equation: 
	\eqs{\frac{r_k}{r_k - a} = \exp( a/f_k)}
Consider $v \in [0,R]$ and define $a=r_k(1 - v/R)$. Substituting in the equation above we get: $R e^{-R} = v e^{-v}$. Since $v \mapsto  v e^{-v}$ is strictly increasing on $[0,1]$ and strictly decreasing on $[1,+\infty)$ this equation has at most two solutions. $v = R$ is a trivial solution, which gives $a = 0$. The other solution is found by noticing that $-R e^{-R} = -v e^{-v}$, so that $-v = W( -R e^{-R})$, and $a =  r_k(1 + W( -R e^{-R})/R)$ which proves the result.
	
	(ii) If $X_k(\ell) \leq 1/f_k$ a.s. then $a^\star_k = +\infty$ so that the claim is trivially true. If we do not have $X_k(\ell) \leq 1/f_k$ a.s., then $F_k(a) \tto{a \to \infty} \infty$. Define $\overline{a} = 2 \mu_k (R-1) / v_k^2$. By continuity of $F_k$, to prove that  $a^\star_k \geq \overline{a}$ it suffices to prove that $ F_k( \overline{a}) \leq 0$. If $X_k(\ell)$ is $v^2_k$-sub-Gaussian, we have $$F_k(a) \leq a \mu_k  +  \frac{a^2 v^2_k}{2} - \frac{a}{f_k} = a \mu_k(1 - R) + \frac{a^2 v^2_k}{2}.$$ Setting $a \equiv \overline{a}$ we obtain that $F_k(\overline{a}) \leq 0$ which concludes the proof. \ep

\begin{remark}\label{rem:subgaussian}
	(a) If $X_k(\ell) \sim {\cal N}(\mu_k,\sigma^2_k)$ then $G_k(a) = a \mu_j + a^2 \sigma_k^2/2$ and $X_k(\ell)$ is $\sigma_k^2$ sub-Gaussian.
	
	(b) If $X_k(\ell) \in [\underline{x},\overline{x}]$ a.s. then $X_k(\ell)$ is $(\overline{x} - \underline{x} )^2/4$ sub-Gaussian (by Hoeffding's lemma).
\end{remark}
 
\begin{theorem}\label{th:iid_upper}
	Consider $\pi$ an $f$-upper balanced policy. Let assumptions~\ref{as:iid} and~\ref{as:links} hold. Define $N_k = f_k N$ and ${\cal G}_k = \{a: F_k(a) \geq 0 \}$.
	
	(i) For all $b \geq 0$  and all $N$ we have:\eqs{
P^{N}(\pi,b+ K - 1) \leq 1 - \prod_{k=1}^K \left[1 - \min_{a_k \in {\cal G}_k} e^{ N_k F_k(a_k) - a_k b} \right].
}
	(ii) If $R > 1$, we have $a_k^\star > 0$ and for all $N$ and all $b > 0$: 
	\eqs{
	  P^{N}(\pi,b + K - 1) \leq 1 - \prod_{k=1}^K [1 - e^{-a_k^\star b}].
	}
	(iii) If $R \leq 1$, and $X_k(\ell)$ is $v_k^2$-sub-Gaussian for all $k$, then for all $b > 0$ and all $N$:
	\eqs{
		P^{N}(\pi, (R^{-1}-1)N + b + K - 1) \leq 1 - \prod_{k=1}^K [1 - e^{ -\frac{b^2}{2 v_k^2 N f_k}}].
	}	
\end{theorem}

\begin{corollary}\label{cor:iid_upper}
	The following holds without assumption~\ref{as:links}. 
		
	(i) For all $b \geq 0$  and all $N$ we have:\eqs{
P^{N}(\pi,b+ K - 1) \leq \sum_{k=1}^K \min_{a_k \in {\cal G}_k} e^{ N_k F_k(a_k) - a_k b}.
}
	(ii) If $R > 1$, we have $a_k^\star > 0$ and for all $N$ and all $b > 0$: 
	\eqs{
	  P^{N}(\pi,b + K - 1) \leq \sum_{k=1}^K e^{-a_k^\star b}.
	}
	(iii) If $R \leq 1$, and $X_k(\ell)$ is $v_k^2$-sub-Gaussian for all $k$, then for all $b > 0$ and all $N$:
	\eqs{
		P^{N}(\pi, (R^{-1}-1)N + b + K - 1) \leq \sum_{k=1}^K e^{-\frac{b^2}{2 v_k^2 N f_k}}.
	}	
\end{corollary}

\bp
(i) Consider $n$ such that $\pi_n = k$. Define $M_k(t) = \sum_{\ell=1}^{t} X_k(\ell)$. Consider $d= d_k(n)$, so that $n$ is the $d$-th chunk requested on link $k$. The pre-buffering time is equal to $b + K - 1$, and chunk $n$ is not received in time if and only if $M_k(d) \geq n + b + K - 1$. Further, since $\pi$ is $f$ upper balanced we have $d = d_k(n) \leq (n + K - 1)f_k$ so that $d/f_k + 1 - K \leq n$. It is also noted that $d_k(n) \leq N_k$. Hence, using independence:
\als{
P^{N}(\pi,b +  K - 1) &\le \PP\left[\exists k : \max_{1 \leq d \leq N_k} M_k(d) - d/f_k \geq b \right] \\
&\le 1 - \prod_{k=1}^K \left(1 - \PP\left[ \max_{1 \leq d \leq N_k} M_k(d) - d/f_k \geq b \right]\right)
}
Consider $a \in {\cal G}_k$. Define $Z_k(d) = e^{a(M_k(d) - d/f_k)}$. $Z_k(d)$ is a sub-martingale, indeed, 
\eqs{\EE[Z_k(d) | Z_k(d-1) ] = Z_k(d-1) e^{ F_k(a)} \geq Z_k(d-1),}
since $a \in {\cal G}_k$. Using Doob's maximal inequality:
\als{
\PP[ \max_{1 \leq d \leq N_k}   M_k(d) - d/f_k \geq b ] &=\PP[ \max_{1 \leq d \leq N_k} Z_k(d) \geq e^{a b} ]
													\leq e^{-ab} \EE[ Z_k(N_k) ]
													= e^{  N_k F_k(a)  - ab} \\
													&\leq \min_{a \in {\cal G}_k}  e^{ N_k F_k(a) - a b}.
}
Replacing we get the first claim:
\als{
P^{N}(\pi,b + K - 1) 
\leq 1 - \prod_{k=1}^K \left[1 - \min_{a_k \in {\cal G}_k} e^{N_k F_k(a_k) - a_k b} \right].
}

(ii) Assume that $R > 1$. For all $k$, we prove that either $X_k(\ell) \leq 1/f_k$ a.s., or otherwise there exists $a_k^\star > 0$ such that $F(a_k^\star) = 0$. The derivative of $F_k(.)$ evaluated at $a = 0$ is $\EE[X_k(\ell)] - 1/f_k = \mu_k(1 - R) < 0$. Therefore $F_k(a) < 0$ for $a > 0$ on an open neighborhood of $0$. Choose $\epsilon > 0 $ such that $\PP[ X_k(\ell) \geq 1/f_k + \epsilon] > 0$. We have:
\als{
F_k(a) \geq \log( \PP[ X_k(\ell) \geq 1/f_k + \epsilon] e^{a (1/f_k + \epsilon)} ) - a /f_k
					\sim a\epsilon \tto{a \to +\infty} +\infty  
}
Hence by continuity of $F(.)$ there exists $a_k^\star > 0$ such that $F(a_k^\star) = 0$. The second claim is obtained by setting $a_k \equiv a^\star_k$ for all $k$ in the first claim.

(iii) Consider $R \leq 1$. Define $B = b + N(R^{-1}-1)$. For all $a \geq 0$, from Jensen's inequality we have $G_k(a) \geq a \mu_k$, so $F_k(a) \geq a \mu_k(1 - R) \geq 0$.  Hence $[0,+\infty) \subset {\cal G}_k$ and $a_k \in {\cal G}_k$. Since $X_k(\ell)$ is $v^2_k$-sub-Gaussian, $G_k(a) \leq a \mu_k + a^2 v_k^2/2$, so that: 
\als{
N_k F_k(a) - a B \leq a( N(R^{-1}-1) - B) + a^2 v_k^2 N f_k /2
								= - a b + a^2 v_k^2 N f_k /2.
}
Setting $a = {b \over v_k^2 N f_k}$ we obtain $N_k F_k(a) - a B \leq - \frac{b^2}{2 v_k^2 N f_k}$. Substituting in the first claim (where $b$ is replaced by $B$) we obtain the announced result.

The corollary follows from the same reasoning but using the union bound (which does not require assumption \ref{as:links} to hold):
\als{
P^{N}(\pi,b +  K - 1) \le \PP\left[\exists k : \max_{1 \leq d \leq N_k} M_k(d) - d/f_k \geq b \right] 
\le \sum_{k=1}^K \PP\left[\max_{1 \leq d \leq N_k} M_k(d) - d/f_k \geq b \right].
}

\ep

\subsection{Performance for Markovian delays}
We turn to Markovian delays. Even for very simple transition matrices $Q^k$ and rate functions $r(.)$ it is difficult to find a tractable formula for the starvation probability, as it involves upper bounding the crossing probability of a random walk with correlated increments (see for instance \cite{Kugler13}). It is noted that the starvation probability does not only depend on stationary distributions of the link states $m_k$ but also on the the transition rate matrix $Q^k$. One can expect larger starvation probability when the Markov chain has a strong correlation since in that case the streaming flow can experience long bursts of large delays. We propose to consider a regime where $S_k(t)$ evolves on a ``faster time scale'' than the streaming flow of interest. In that case we can find a simple, tractable expression for the starvation probability shown in theorem~\ref{th:markov_upper}. This approximation is useful in practice, since streaming flows are long (several minutes), while most phenomena causing link variability are short, for instance fading, medium access protocols and short lived elastic flows.  We consider speed $\phi > 0$, and we define the accelerated process $( S_k( \phi t) )_{t}$. This process is again a continuous time, stationary ergodic Markov chain, with the same stationary distribution $m_k$, and transition rate matrix $\phi Q^k$. 

 Theorem~\ref{th:markov_upper}, statement (i) shows the intuitive fact that when $R > 1$ and one lets $\phi \to \infty$, the starvation probability vanishes. Indeed, seen from the streaming flow, the link variability disappears and the chunk delays become equal (almost surely) to $r_k$ due to the ergodic theorem for Markov chains,  This suggests to look at a setting when $R$ depends on $\phi$ and approaches $1$ as $\phi \to \infty$. To avoid a trivial result (i.e. $P \to 0$ or $P \to 1$ when $\phi \to \infty$), the sum rate and pre-buffering time should scale as $1/(1 - C_1/\sqrt{\phi})$ and $C_2/\sqrt{\phi} + K - 1$ respectively with $C_1,C_2$ two positive constants, as done in theorem~\ref{th:markov_upper}, statement (ii). Before stating our result we state lemma~\ref{lem:battacharya} due to~\cite{Bhattacharya82}, which shows that the amount of data received on a link can be approximated by a Wiener process with appropriate drift and variance. By a slight abuse of notation, we identify $r(.)$ with the vector $(r(i))_{i \in {\cal S}}$.
\begin{lemma}[\cite{Bhattacharya82}]\label{lem:battacharya}
	Define $G^{\phi}(t) = \sqrt{\phi} \int_{0}^t (r( S_k(\phi u))- r_k)du$. Consider $g^k = (g^k(i))_{i \in {\cal S}}$ a solution to the Poisson equation: $Q^k g^k  = r(.) - r_k$. Further define the asymptotic variance : 
	\eqs{
	\overline{\sigma}_k^2 =  - 2 \sum_{i \in {\cal S}} r(i) g^k(i) m_k(i).
	}
	Then $G^{\phi}(.)$ converges to a Wiener process with drift $0$ and variance $\overline{\sigma}_k^2$, when $\phi \to \infty$.
\end{lemma}
 
Given transition matrix $Q^k$, the Poisson equation can be solved numerically by calculating the pseudo inverse of $Q^k$. We show below that for in several cases of interest, one can solve the Poisson equation in closed form and deduce the asymptotic variance $\overline{\sigma}_k^2$. This makes our performance bounds fully tractable. We may now state theorem~\ref{th:markov_upper}. The proof uses standard results on crossing probabilities for Wiener processes, and the interested reader can refer to \cite{Resnick,Billingsley} (for instance).
\begin{theorem}\label{th:markov_upper}
Consider $\pi$ an $f$-upper balanced policy. Let assumptions~\ref{as:markov} and \ref{as:links} hold.

(i) Consider $R > 1$ and $b > 0$ fixed. Then for all $N$:
\eqs{
	 P^{N,\phi}(\pi, b + K - 1) \tto{\phi \to \infty} 0.
}

(ii) Consider $C_1,C_2 \geq 0$, define $b^{\phi} = C_2/\sqrt{\phi}$ and assume that $R = R^{\phi} \equiv 1/(1 - C_1/\sqrt{\phi})$. Then for all $N$: 
\als{
	\lim \sup_{\phi \to \infty} P^{N,\phi}(\pi, b^{\phi} + K - 1) 
	\leq 1 - \prod_{k=1}^{K} \left( 1 - \PP\left[ \max_{t \in \lbracket K,N+K-1\rbracket} \hspace{-0.5cm} \overline{\sigma}_k W(t) - r_k C_1 t \geq r_k C_2 \right] \right).	
}
with $(W(t))_t$ a standard Wiener process.

(iii)(a) If $C_1 = 0$, we have:
\als{
\lim \sup_{\phi \to \infty} P^{N,\phi}(\pi, b^{\phi} + K - 1)  
\leq 1 - \prod_{k=1}^{K} \left( 1 -  2 \Psi\left( \frac{r_k C_2}{\overline{\sigma}_k \sqrt{N + K - 1}}\right) \right)
}
(iii)(b) If $C_1 > 0$, for all $N$:
 \eqs{
\lim \sup_{\phi \to \infty} P^{N,\phi}(\pi, b^{\phi} + K - 1) \leq 1 - \prod_{k=1}^{K} \left( 1 - e^{ - 2 \frac{r_k^2 C_1 C_2}{\overline{\sigma}_k^2}} \right).
}
\end{theorem}

\begin{corollary}\label{cor:markov_upper}
 The following holds without assumption~\ref{as:links}.

(i) Consider $R > 1$ and $b > 0$ fixed. Then for all $N$:
\eqs{
	 P^{N,\phi}(\pi, b + K - 1) \tto{\phi \to \infty} 0.
}

(ii) Let $C_1,C_2 \geq 0$, $b^{\phi} = C_2/\sqrt{\phi}$ and $R = R^{\phi} \equiv 1/(1 - C_1/\sqrt{\phi})$. For all $N$:
\als{
	\lim \sup_{\phi \to \infty} P^{N,\phi}(\pi, b^{\phi} + K - 1) 
	\leq \sum_{k=1}^{K} \PP\left[ \max_{t \in \lbracket K,N+K-1\rbracket} \overline{\sigma}_k W(t) - r_k C_1 t \geq r_k C_2 \right].	
}
with $(W(t))_t$ a standard Wiener process.

(iii)(a) If $C_1 = 0$, we have:
\als{
\lim \sup_{\phi \to \infty} P^{N,\phi}(\pi, b^{\phi} + K - 1) \leq \sum_{k=1}^K 2 \Psi\left( \frac{r_k C_2}{\overline{\sigma}_k \sqrt{N + K - 1}}\right)
}
(iii)(b) If $C_1 > 0$, for all $N$:
 \eqs{
\lim \sup_{\phi \to \infty} P^{N,\phi}(\pi, b^{\phi} + K - 1) \leq \sum_{k=1}^{K}  e^{ - 2 \frac{r_k^2 C_1 C_2}{\overline{\sigma}_k^2}}.
}
\end{corollary}

\bp

	(i) We omit superscripts $.^{\phi}$ for clarity. We define $H_k(t) = \int_{0}^t r(S_k(\phi u)) du$, and $G_k(t)= \sqrt{\phi} (H_k(t) - r_k t)$ and $B = b + K - 1$. Consider chunk $n$, to be read at time $n + B$, and requested on link $k$, so that $\pi_n = k$. Chunk $n$ does not arrive in time if and only if less than $d_k(n)$ chunks have been received on link $k$ at time $n + B$, so that $H_k(n + B) \leq d_k(n) \leq f_k( n + K - 1)$ since $\pi$ is $f$-upper balanced. Hence $H_k(n + B) - (n + B) f_k  \leq - f_k b$. Therefore the starvation probability can be upper bounded by:
	\eqs{
	 \PP[ \exists k: \min_{n \in \lbracket 1,N\rbracket}  H_k(n + B) - (n + B) f_k  \leq - f_k b  ].
	}
	Now, from lemma~\ref{lem:battacharya}, we know that, for all $n$,
	\als{
	H_k(n + B) - (n + B)f_k \tto{\phi \to \infty} (n + B) (r_k - f_k) 
	= (n + B) r_k(1 - 1/R) > 0,
	} since $R > 1$. We deduce that the starvation probability goes to $0$ when $\phi \to \infty$ as announced.
	
	(ii) It is noted that $R$, $f_k$ and $B = b^{\phi} + K - 1$ depend on $\phi$, and we omit the superscript $.^{\phi}$ for clarity. Assume that event $\{H_k(n + B) - (n + B) f_k  \leq - f_k b\}$ occurs. Then:
	\eqs{
		G_k(n + B) + \sqrt{\phi}(r_k - f_k)(n + B) \leq - \sqrt{\phi} f_k b
	}
	We have assumed that $f_k = r_k/R = r_k( 1 - C_1/\sqrt{\phi})$ and $b = C_2/\sqrt{\phi}$, so that replacing above yields:
	\eqs{
		G_k(n + B) + r_k C_1 (n + B) \leq - C_2 r_k( 1 - C_1/\sqrt{\phi}). 
	}
	So the starvation probability is upper bounded by:
	\als{
	 \overline{P} &\equiv \PP[ \exists k: \hspace{-0.3cm} \min_{t \in \lbracket K,N + K - 1\rbracket} \hspace{-0.3cm} G_k(t) + t r_k C_1  \leq - C_2 r_k( 1 - C_1/\sqrt{\phi})].
	}
	Letting $\phi \to \infty$ and using lemma~\ref{lem:battacharya}:
	\als{
	 	\lim \sup_{\phi \to \infty} \overline{P} &\le \PP[ \exists k: \hspace{-0.3cm} \max_{t \in \lbracket K,N + K - 1\rbracket}  \overline{\sigma}_k W_k(t) - t r_k  C_1  \geq r_k C_2  ] \sk
	 	&= 1 - \prod_{k=1}^K \left( 1-  \PP\left[ \max_{t \in \lbracket K,N + K - 1\rbracket} \hspace{-0.5cm} \overline{\sigma}_k W(t) - t r_k C_1  \geq r_k C_2\right]\right).
	}
	where $W_1(.),...,W_K(.)$ are independent standard Wiener processes, and where we used independence to obtain the last equality.	
	
	(iii)(a) Consider $C_1 = 0$. We recall the reflexion principle. 
	\begin{proposition}
	Consider $W(t)$ a standard Brownian motion, then for all $\omega > 0$ and $T \geq 0$:
	$$\PP\left( \sup_{0 \leq t \leq T} W(t) \geq \omega\right) = 2 \PP(W(T) \geq \omega).$$   
	\end{proposition}
	
	Now using the reflexion principle:
	\als{
	\PP  \left[ \max_{t \in \lbracket K,N + K - 1\rbracket} \overline{\sigma}_k W(t)  \geq r_k C_2\right] &\leq \PP  \left[ \max_{t \in [0, N + K - 1]} \overline{\sigma}_k W(t)  \geq r_k C_2\right] \sk
	&= 2 \PP\left[  W(N + K - 1)  \geq r_k C_2/ \overline{\sigma}_k\right] \sk
	&= 2 \Psi\left( \frac{r_k C_2}{\overline{\sigma}_k \sqrt{N + K - 1}}\right),
	}
	and replacing in (ii) gives the announced statement.
	
	(iii)(b) Consider $C_1 > 0$. We first recall a result on the probability that a Wiener process with negative drift hits a positive level. 
	\begin{proposition}
	Consider $W(t)$ a standard Brownian motion, $\sigma > 0$, $\mu > 0$ and $\omega > 0$. Then:
	$$\PP\left( \sup_{t \geq 0}( \sigma W(t) - \mu t) \geq \omega\right) = e^{- {2 \mu \omega \over \sigma^2}}.$$  
	\end{proposition}
	We apply the result above to yield: 
	\als{
	\PP \left[ \max_{t \in \lbracket K,N+K-1\rbracket} \hspace{-0.3cm} (\overline{\sigma}_k W(t) - r_k C_1 t) \geq r_k C_2 \right] 
	\leq \PP\left[ \sup_{ t \geq 0} (\overline{\sigma}_k W(t) - r_k C_1 t)  \geq r_k C_2\right] 
	= e^{ - 2 \frac{r_k^2 C_1 C_2}{\overline{\sigma}_k^2}}. 
	}
	Replacing in (ii) gives the announced result. 
	
	The corollary is obtained by the same reasoning using a union bound (which does not require assumption~\ref{as:links}):
\als{
	 \PP[ \exists k: \min_{n \in \lbracket 1,N\rbracket}  H_k(n + B) - (n + B) f_k  \leq - f_k b  ] 
	 \le  \sum_{k=1}^K \PP[ \min_{n \in \lbracket 1,N\rbracket}  H_k(n + B) - (n + B) f_k  \leq - f_k b].
	}
	\ep
	
The result is also valid when $R = 1$ (critical case) since $C_1$ can be null. If $C_1 = C_2$ the bound still holds but is not informative.
\section{Case Studies}\label{sec:case_studies}
In this section we specialize our results to several simple models for wired and wireless links. We show how to calculate relevant quantities such as $G_k(.)$ and $\overline{\sigma}^2$. We focus on a particular link and indexes $._{k}$ , $.^{k}$ are omitted for clarity. We consider simple models for tractability. It is noted that as far as CSMA and opportunistic scheduling are concerned, one could consider rate adaptation (at the expense of more complicated expressions) and still derive the cumulant generating function, since for reasonable models, delays follow a phase-type distribution.
\subsection{Wireless links with random access (CSMA-like)}\label{subsec:wifi}
We consider the following model for random access, which is Bianchi's model~\cite{Bianchi00} with one back-off stage. A fixed number of flows compete for access to a link. Each chunk is split into $n_f$ frames. Time is slotted, with $T_s$ the time slot duration, and when the streaming flow attempts to access the link, it is successful with probability $p$. If the attempt is successful, a frame is transmitted during a slot. If the attempt is unsucessful, then the streaming flow waits for a duration uniformly distributed between $0$ and $W T_s$, with $W$ the window size. Given the number of competing flows and the window size, $p$ may be calculated using a fixed point equation as in~\cite{Bianchi00}. The time required to transmit a frame is $Z = T_s(1 + W \sum_{i=1}^G U_i) $ with $G \sim$Geo($p$) and $(U_{i})$ i.i.d. uniform on $[0,1]$ and independent of $G$. Define $a' = a T_s W$. We have $\EE[e^{a' U_i}] = \int_0^{1}  e^{a' u} du = h(a')$. $(U_i)_i$ are i.i.d and independent from $G$ so:
\als{
	\EE[ e^{a' \sum_{i=1}^G U_i}] &= \sum_{g \ge 0} \PP[ G = g ] \left( \prod_{i=1}^g \EE[e^{a' U_i}] \right)
	=  \sum_{g \ge 0} p(1-p)^g h(a')^g = {p \over 1 - (1-p)h(a')}, 
}
if $(1-p) h(a') < 1$ and $=\infty$ otherwise. We deduce:
$$
	\log( \EE[ e^{a Z} ]) = T_s a +  \log \left( \frac{p}{1 - (1-p)h(a W T_s)} \right).
$$
The time to transmit a chunk is the sum of $n_f$ i.i.d copies of $Z$, and we deduce the cumulant generating function:
\als{
	G(a) =  n_f \left( T_s a + \log \left( \frac{p}{1 - (1-p)h(a W T_s)} \right) \right) \,\,\, \text{ , } \,\,\,
	 h(a)=  (e^a-1)/a.
}
with $a$ such that $(1-p) h(a W T_s) < 1$.
\subsection{Wireless links using channel-aware scheduling}\label{subsec:cellular}
We consider channel-aware scheduling with a large number of flows. Once again chunks are divided in $n_f$ frames. In each time slot (duration $T_s$), a scheduler chooses the flow whose ratio between instantaneous data rate and expected data rate is maximal and the chosen flow transmits a frame. Define $p$ the inverse of the number of competing flows, the time required to transmit a frame is $Z = T_s(1 + G)$ where $G \sim$ Geo($p$), and the time required to transmit a chunk is a sum of $n_f$ i.i.d copies of $Z$. We deduce the cumulant generating function: \eqs{
G(a) = n_f  \left( a T_s + \log \left( \frac{p}{1 - (1-p)e^{a T_s}} \right) \right).
}
with $a < -\log(1-p)/T_s$. This model is reasonable if the number of flows is large, so that the instantaneous data rate of the chosen user equals the peak rate (the rate achieved with maximal modulation and coding scheme).

\subsection{Wireless ON-OFF channels}\label{subsec:onoff}

 We consider a typical model used in cognitive radio. A link is shared between a primary user and the streaming flow acting as a secondary user. The primary user's activity follows a two-states Markov process, which is independent of the secondary user activity. The streaming flow transmits only when the primary user is not transmitting. The link has two states: state $0$ when the primary user is active and no data is transmitted so that $r(0) = 0$ and state $1$, when the link is available and data is transmitted at rate $r(1)=1$. The transition rate matrix is $Q = \begin{pmatrix} -\beta & \beta \\ \alpha & -\alpha \end{pmatrix}$. The stationary distribution is $m = ( \frac{\alpha}{\alpha + \beta},\frac{\beta}{\alpha + \beta} )$ and the expected data rate is $r = \frac{\beta}{\alpha + \beta}$. The Poisson equation reads $(-r,1-r) = (g^k_1-g^k_0) ( \beta, - \alpha)$ and a solution is: $g^k = ( 0 , \frac{-1}{\alpha + \beta})$. The asymptotic variance is:  
\eqs{
\overline{\sigma}^2 = \frac{2 \beta \alpha}{(\alpha + \beta)^3}.
}
As expected, $\overline{\sigma}^2$ is large when $\alpha + \beta$ is small, since in that case the channel state has a strong time correlation.
\subsection{Sharing links with small flows}\label{subsec:small_flows}
The last model we consider is a link shared between the streaming flow and small exponential flows arriving as a Poisson process. Some form of resource sharing is used (for instance fair rate sharing) and when there are $n$ small flows, the streaming flow transmits data at rate $r(n)$. For fair rate sharing we have $r(n) = 1/(1 + n)$. The state of the link $S(t)$ is the number of small flows at time $t$, and follows an M/M/1 process with arrival rate $\lambda$ and service rate $\mu$. Define the load $\rho = \lambda/\mu < 1$. The stationary distribution is $m(n) = \rho^n(1 - \rho)$. The expected data rate is $\overline{r} = \sum_{n \geq 0} r(n) \rho^n( 1 - \rho)$. Define $\overline{R}(n)= r(n) - \overline{r}$. We now solve the Poisson equation. 
\begin{proposition}
The asymptotic variance $\overline{\sigma}^2$ is:
\als{
\overline{\sigma}^2 = {2 \over \mu} \sum_{n \geq 0} \sum_{i=0}^{n-1} \overline{R}(n) \overline{R}(i) (\rho^{n} - \rho^{i}).
}
\end{proposition}
\bp
By homogeneity, for a fixed value of $\rho$, it sufficient to solve the Poisson equation for $\lambda = 1$, and divide the obtained solution by $\lambda$. For $\lambda = 1$ the Poisson equation reads:
\als{
\overline{R}(0) &=  g(1) - g(0) \sk
\overline{R}(n) &= g(n+1) + \rho^{-1} g(n-1) - (1 + \rho^{-1}) g(n)  \ \ , \ \ n \geq 1.
}
If $g$ is a solution then for all $x \in \RR$, $g' = g + x$ is also a solution, which can be checked by inspection. Hence we look for a solution that verifies $g(0) = 0$. We deduce that $g(1) = \overline{R}(0)$. Define $\tilde{g}(n) = g(n) \rho^n( 1 - \rho)$. Multiplying the previous equation by $\rho^{n+1}( 1 - \rho)$ we get: 
\eqs{
	\tilde{g}(n+1) - \tilde{g}(n) = \rho^{n+1}( 1 - \rho) \overline{R}(n) + \rho(  \tilde{g}(n) - \tilde{g}(n-1) ).
}
One may readily check that $\tilde{g}(n+1) - \tilde{g}(n) = \rho^n (1-\rho) \sum_{k=0}^n \overline{R}(k) $ satisfies the above recursion, and we deduce $\tilde{g}$ by summing its increments: 
\eqs{
\tilde{g}(n) = \tilde{g}(0) + \sum_{i=0}^{n-1} \tilde{g}(i+1) - \tilde{g}(i)
			= \rho \sum_{i=0}^{n-1} \overline{R}(i) ( \rho^{i} - \rho^{n}).
}
Replacing $\tilde{g}$, we obtain the asymptotic variance for $\lambda = 1$:
\als{
\overline{\sigma}^2 = - 2 \sum_{n \geq 0}  \overline{R}(n) \tilde{g}(n) = 2 \rho \sum_{n \geq 0} \sum_{i=0}^{n-1} \overline{R}(n) \overline{R}(i) (\rho^{n} - \rho^{i}).
}
and dividing by $\lambda$ yields the result.\ep
\section{Numerical Experiments}\label{sec:numerical}
We evaluate the numerical performance of the proposed schemes and the accuracy of the various bounds derived above. Throughout this section we consider $N = 3600$ (a video file of 1 hour with chunks of 1 second). We simply simulate the model described in section III, we do not simulate the actual transmission and decoding of chunks. To calculate the starvation probability, we simulate the successive delays on each link $(X_k(\ell))_{k,\ell}$, calculate whether or not starvation occurs and average the result over $10^6$ independent runs.

\subsection{I.i.d. delays}
\begin{figure}
\begin{center}
	\subfigure[Exponential]{\includegraphics[width=\figsizeA]{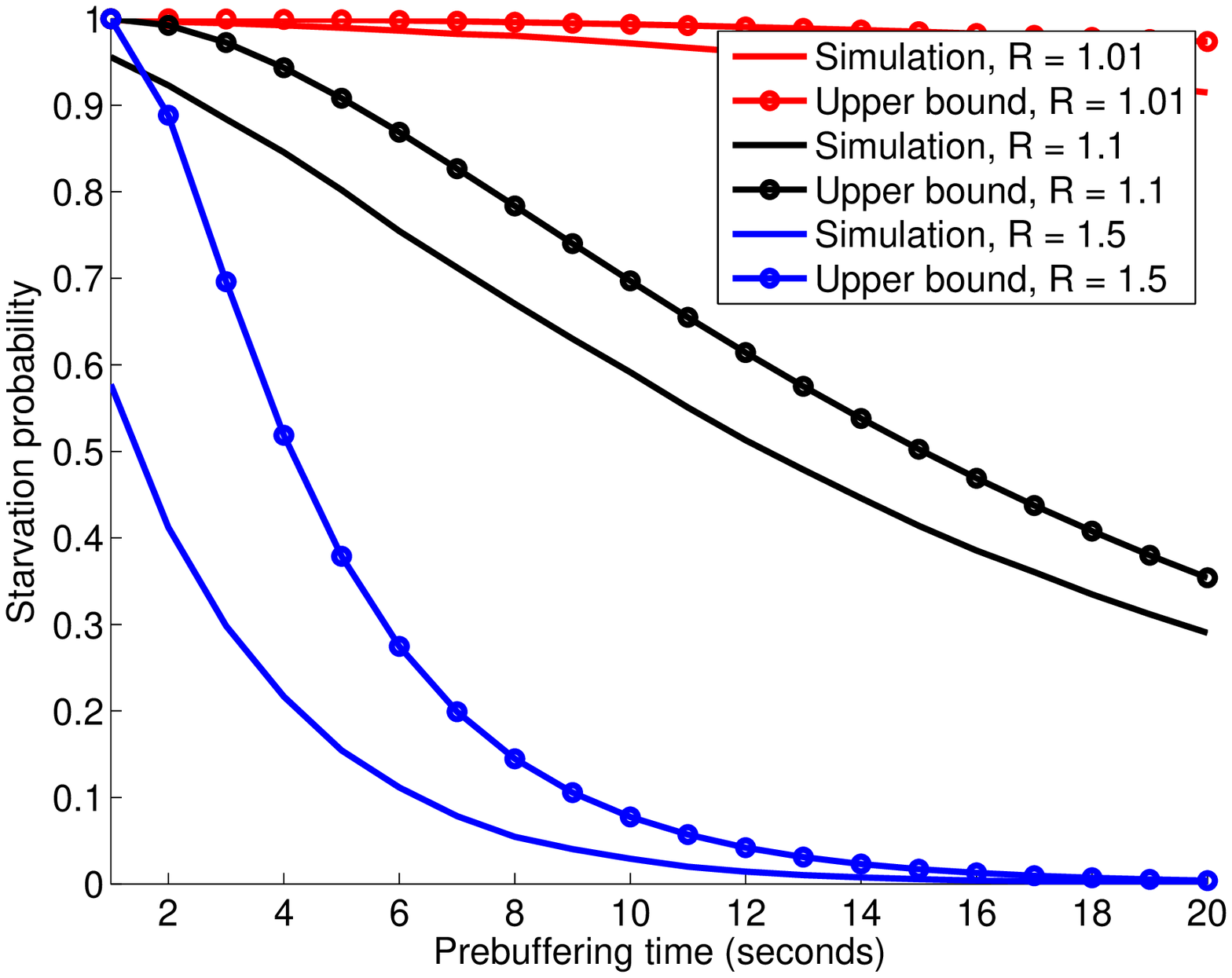}}
    \subfigure[Gaussian]{\includegraphics[width=\figsizeA]{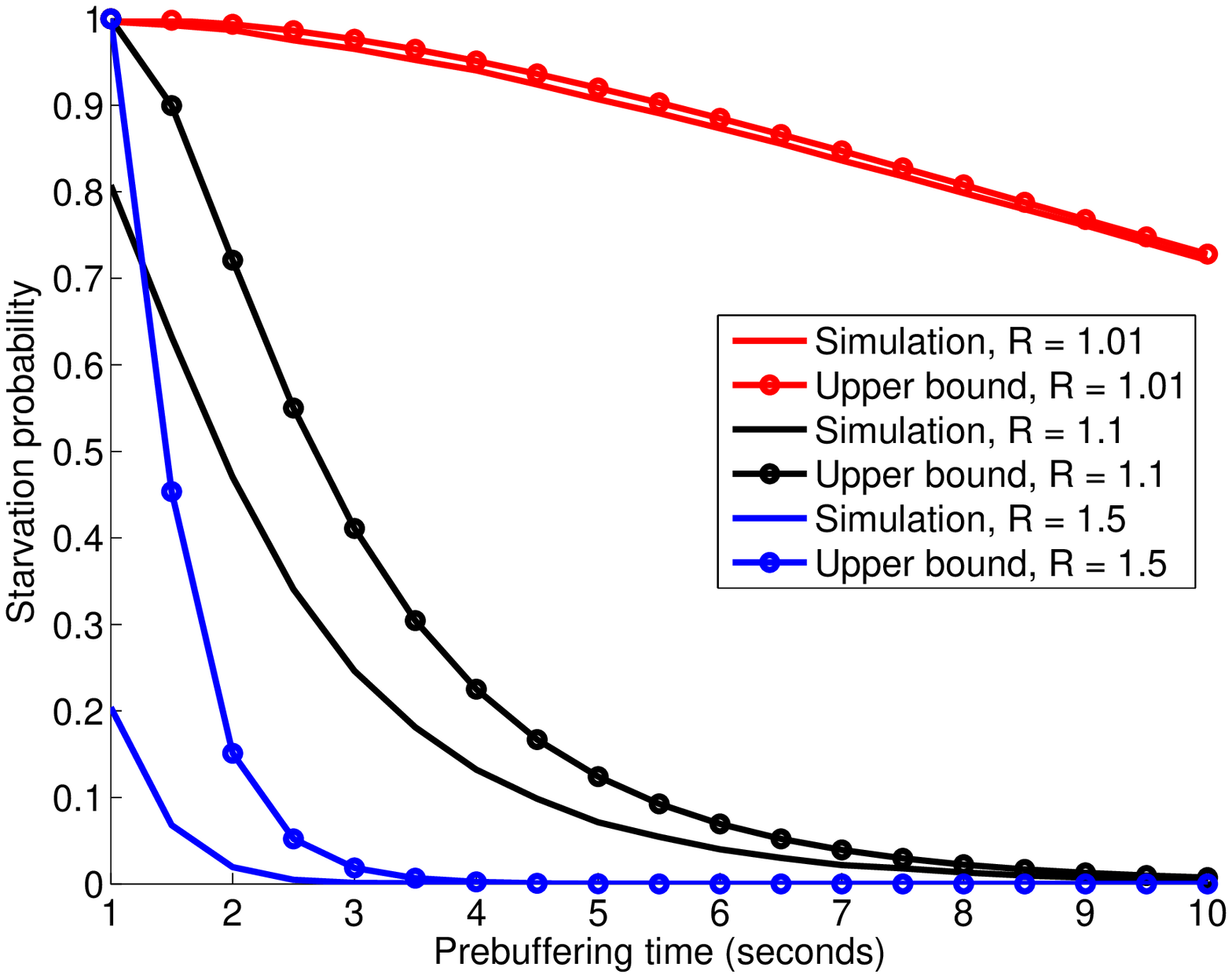}}
\end{center}
\caption{I.i.d. delays, $R > 1$, starvation probability vs upper bound.} \label{fig:iid}
\end{figure}

 We first consider i.i.d delays. On Figure~\ref{fig:iid}(a) we consider two links with i.i.d exponential delays, and we plot the starvation probability as a function of the pre-buffering time, for various values of $R$. Figure~\ref{fig:iid}(b) shows the same for Gaussian delays with variances $0.5$. In both cases our upper bounds are close to the true starvation probabilities, and their accuracy improves when we approach the critical regime $R \approx 1$. The critical regime is the most interesting in practice, since rate adaptation selects a video data rate close to $R$ to ensure maximal quality while avoiding overload.

\begin{figure}[htb]
\begin{center}
\includegraphics[width=\figsize]{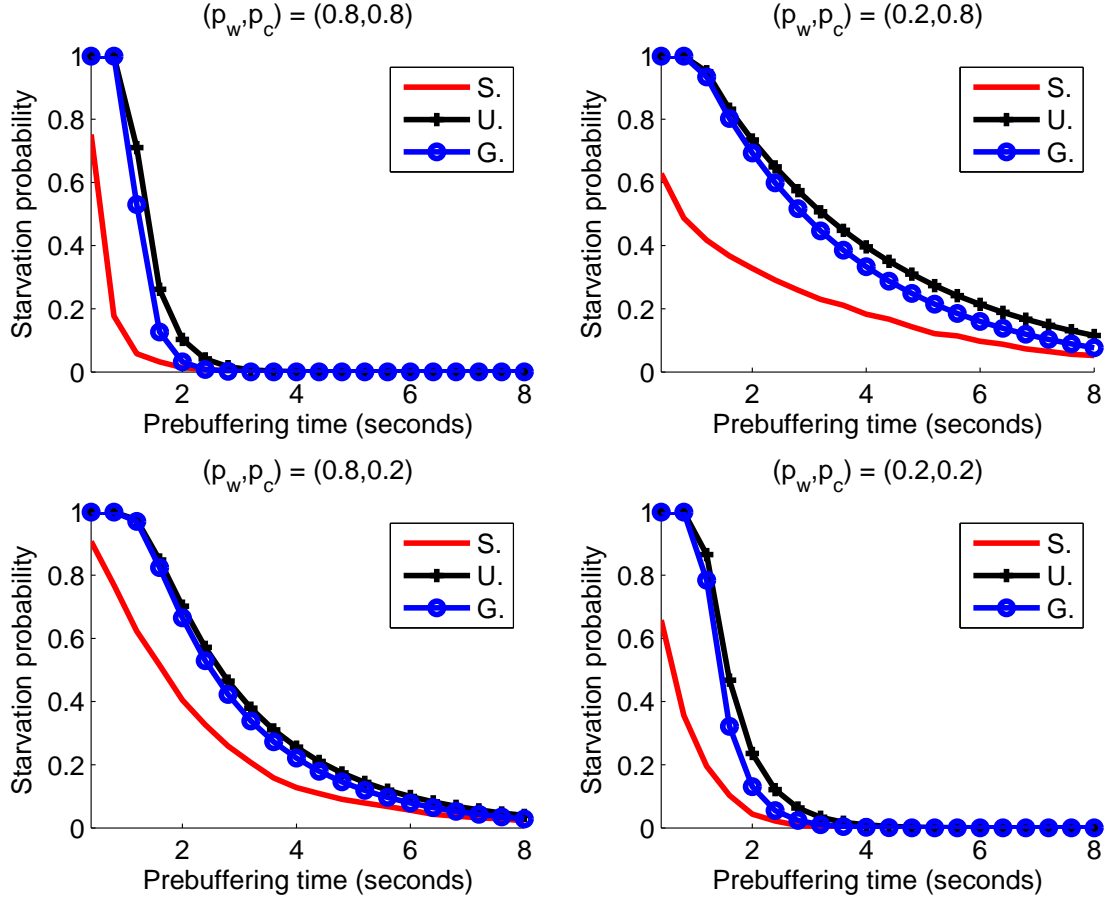} 
\end{center}
\caption{Wifi + cellular links, $R > 1$,  starvation probability vs upper bound.}
\label{fig:wifilte}
\end{figure}

On Figure~\ref{fig:wifilte} we consider two heterogeneous links. The first link is a CSMA-type link (subsection \ref{subsec:wifi}) with access probability $p_w$ and window size of $4$ time slots. The second link uses opportunistic scheduling (subsection~\ref{subsec:cellular}) with access probability $p_c$. For both links, the slot size is $T_s = 10ms$ and each chunk is made of $n_f = 100$ frames. The video data rate is chosen to ensure that we have $R > 1$ and close to $1$. This scenario represents a streaming flow split between cellular and WiFi links. We present the starvation probability as a function of the pre-buffering times calculated by simulation, our upper bound, and an approximation obtained by replacing the delays by Gaussian delays with the same mean and variance. In the legend, ''S.'' stands for simulation, ''U.'' for Upper Bound and ''G.'' for Gaussian Approximation. The Gaussian approximation is fair, due to the fact that the delay of a chunk is the sum of a large number of independent random variables (central limit theorem).
\subsection{Markovian delays}
\begin{figure}
\begin{center}
	\subfigure[ON-OFF links]{\includegraphics[width=\figsizeA]{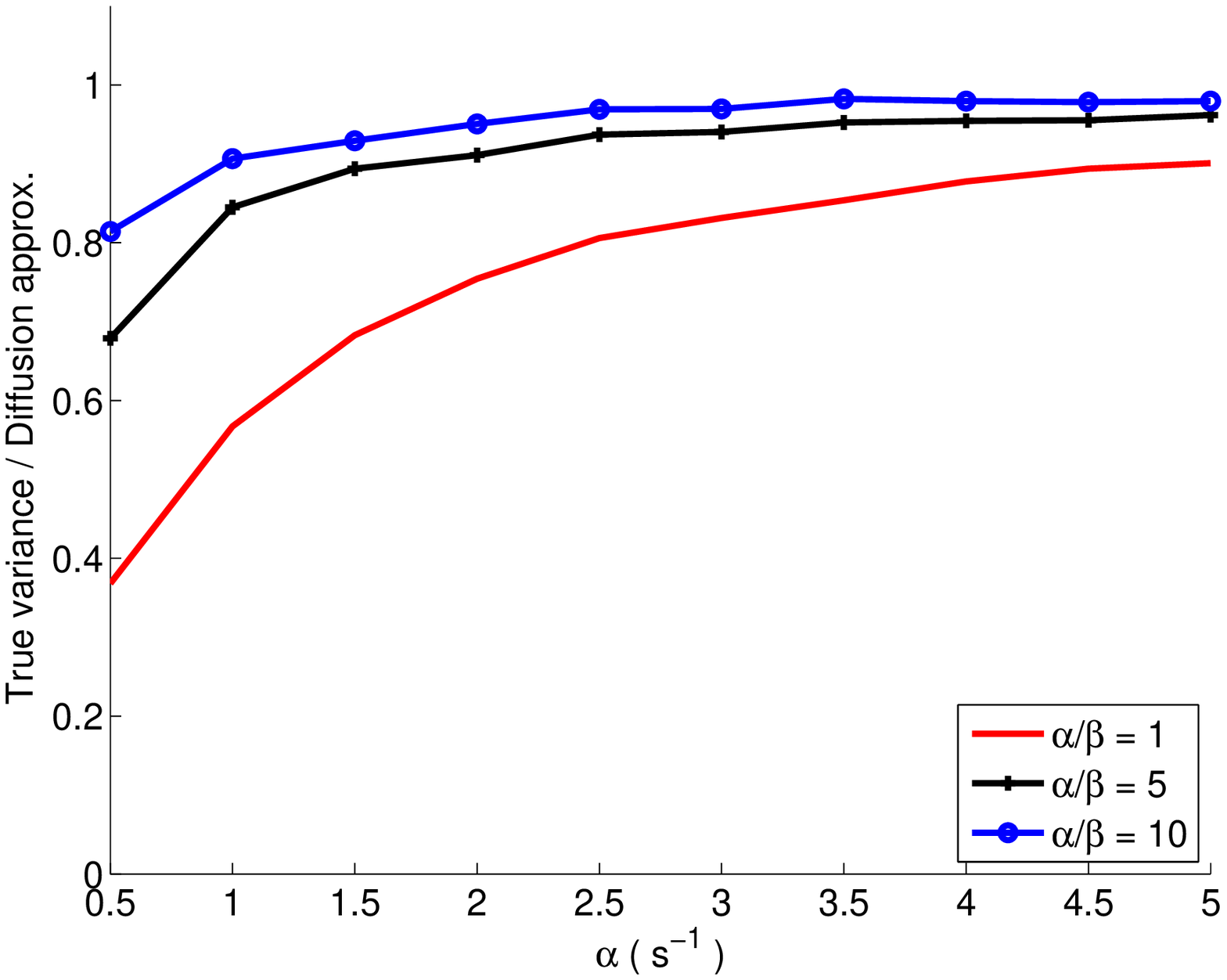}}
    \subfigure[Fair rate sharing]{\includegraphics[width=\figsizeA]{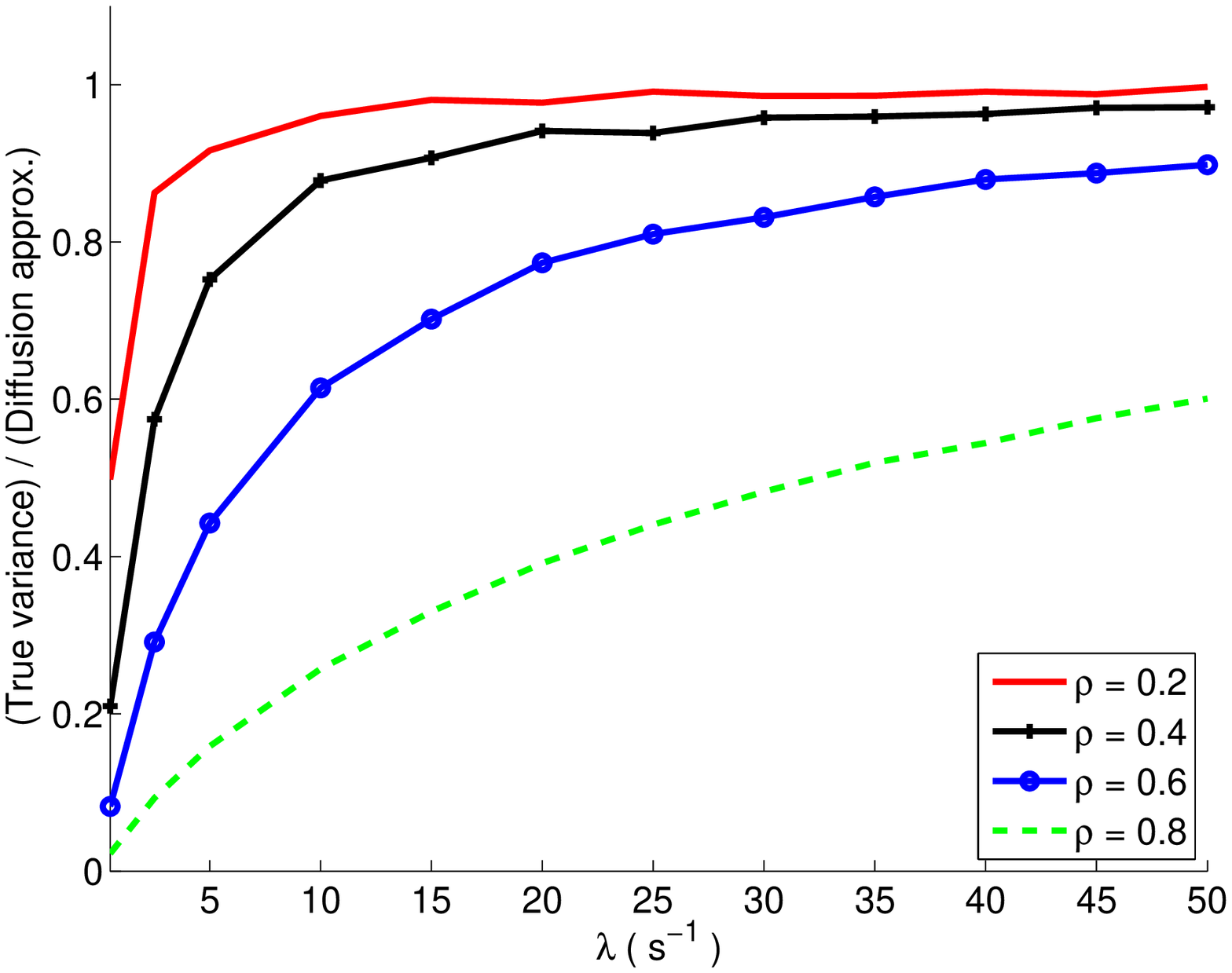}}
\end{center}
\caption{Solution to the Poisson equation vs variance of data rates.} \label{fig:diff}
\end{figure}

We turn to Markovian delays. On Figure~\ref{fig:diff}(a) we consider an ON-OFF link (subsection~\ref{subsec:onoff}). We represent the ratio between the asymptotic variance $\overline{\sigma}_k^2$ and the variance of the amount of data received during a unit of time $\int_{0}^{1} r( S_k(u) ) du$, as a function of the transition rate $\alpha$. As expected, when $\alpha$ grows, the Markov chain $S_k(u)$ moves faster, and we approach the asymptotic regime where $\int_{0}^{1} r( S_k(u) ) du$ becomes normally distributed with mean $r_k$ and variance $\overline{\sigma}_k^2$. Further, this happens for reasonably small values of $\alpha$.

Figure~\ref{fig:diff}(b) shows the same for a link shared with small flows using fair rate sharing (subsection~\ref{subsec:small_flows}). The same conclusions hold, so that the variance of $\int_{0}^{1} r( S_k(u) ) du$ approaches $\overline{\sigma}_k^2$ when $\lambda$ grows (the chain moves faster). Furthermore, we see that the convergence is faster for small loads. This is logical since the mixing time of the chain grows with $\rho$: the process $S_k(t)$ has a stronger time correlation for high loads.

\begin{figure}
\begin{center}
	\subfigure[ON-OFF links]{\includegraphics[width=\figsizeA]{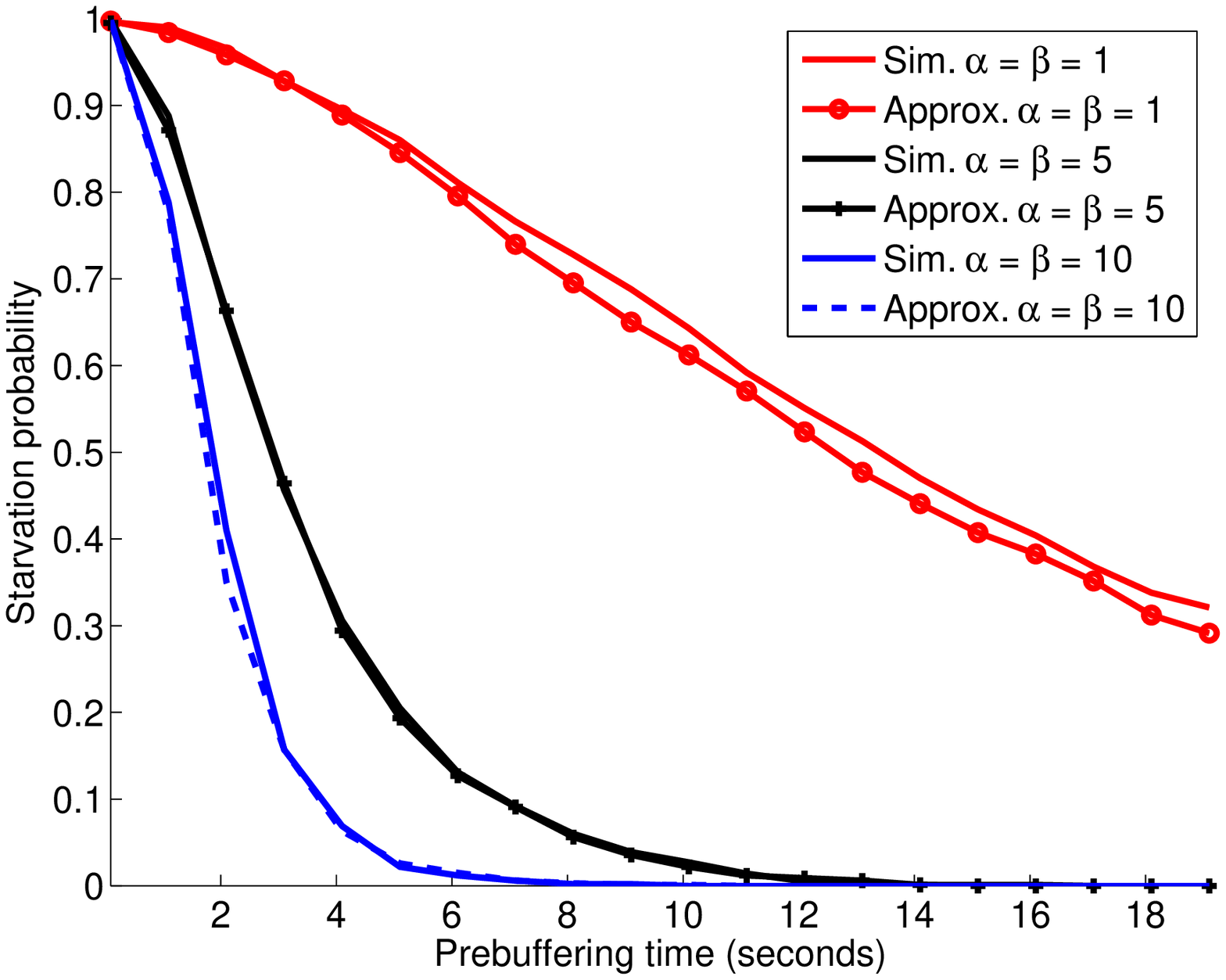}}
    \subfigure[Fair rate sharing, $\rho = 0.7$]{\includegraphics[width=\figsizeA]{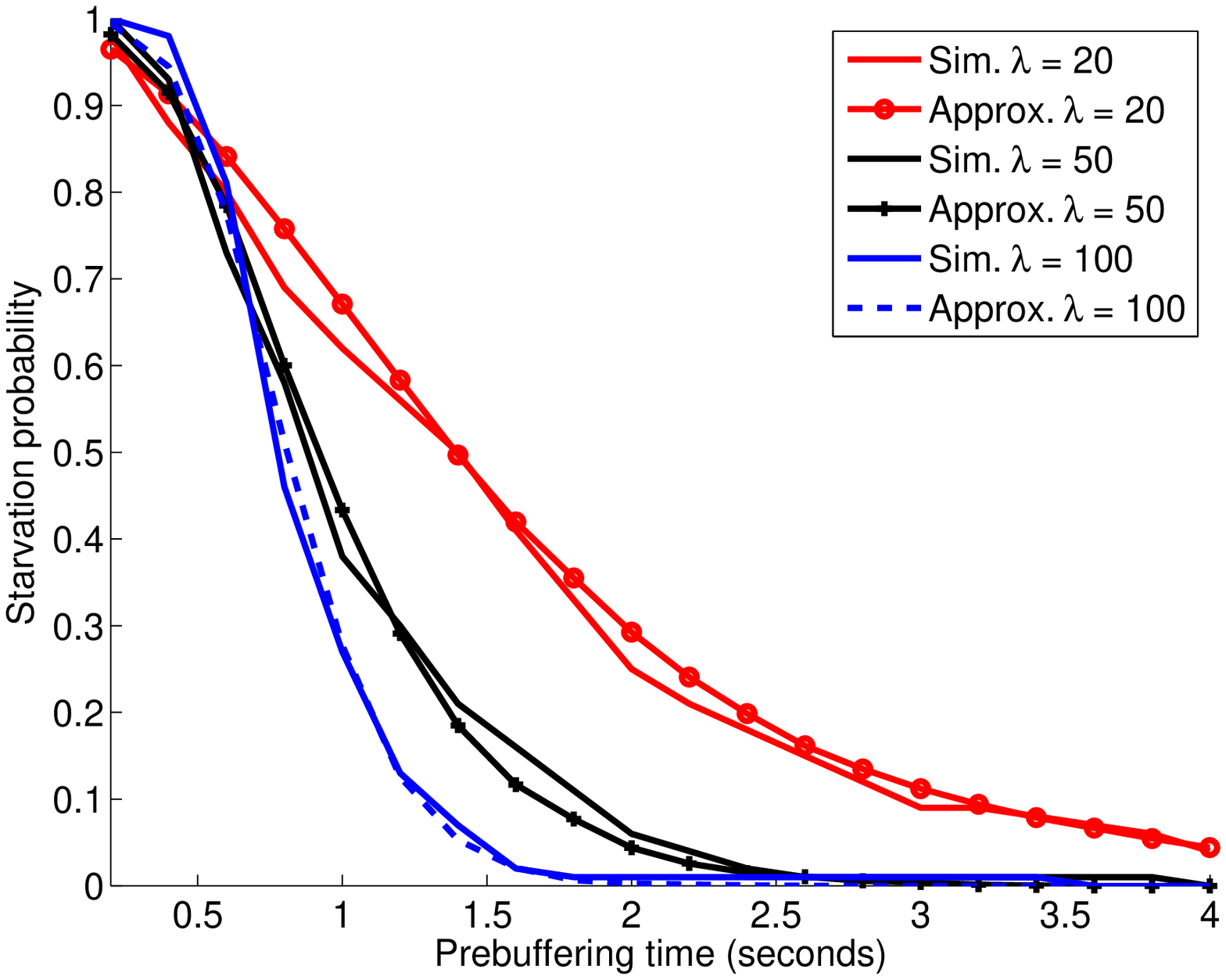}}
\end{center}
\caption{Starvation probability vs diffusion approximation.} \label{fig:starvdiffusion}
\end{figure}

On Figure~\ref{fig:starvdiffusion} we plot the starvation probability as a function of the pre-buffering time for two symmetrical ON-OFF links (resp. links with fair rate sharing and load $\rho = 0.7$). We compare the starvation probability to the diffusion approximation suggested in theorem~\ref{th:markov_upper}. In both cases the diffusion approximation is surprisingly accurate, and gives a tractable approximation to an otherwise intractable (to the best of our knowledge) problem.
\subsection{Real world data}
Finally we present some experiments on real world traces. We store a file of $32$ MB on an HTTP server and, using HTTP byte range requests, we successively request chunks of $100$ kB of this file and measure the corresponding delay. This gives us a trace with $320$ succesive chunk delays. We repeat this $20$ times to obtain $20$ traces. We go through this process twice: first for a wired link (a laptop connected to the Internet through Ethernet, labelled ``Link 1''), and a wireless link (the same laptop connected to the Internet through a WiFi router, labelled ``Link 2''). Links are heterogenous, and we have ${r_1 \over r_2} \approx 2$. 

In table~\ref{tab1} we present the absolute value of the autocorrelation of delays
$
{|\EE[ (X_k(\ell) - \mu_k) (X_k(\ell + L) - \mu_k)]| \over \sigma_k^2}
$
estimated from our traces for both links, where $L$ denotes the lag. Given link $k$ we estimate this quantity and average the result over traces. Delays are only  weakly correlated, so that assuming i.i.d. delays (Assumption~\ref{as:iid}) seems adequate.

\begin{figure}
\begin{center}
	\subfigure{\includegraphics[width=\figsizeA]{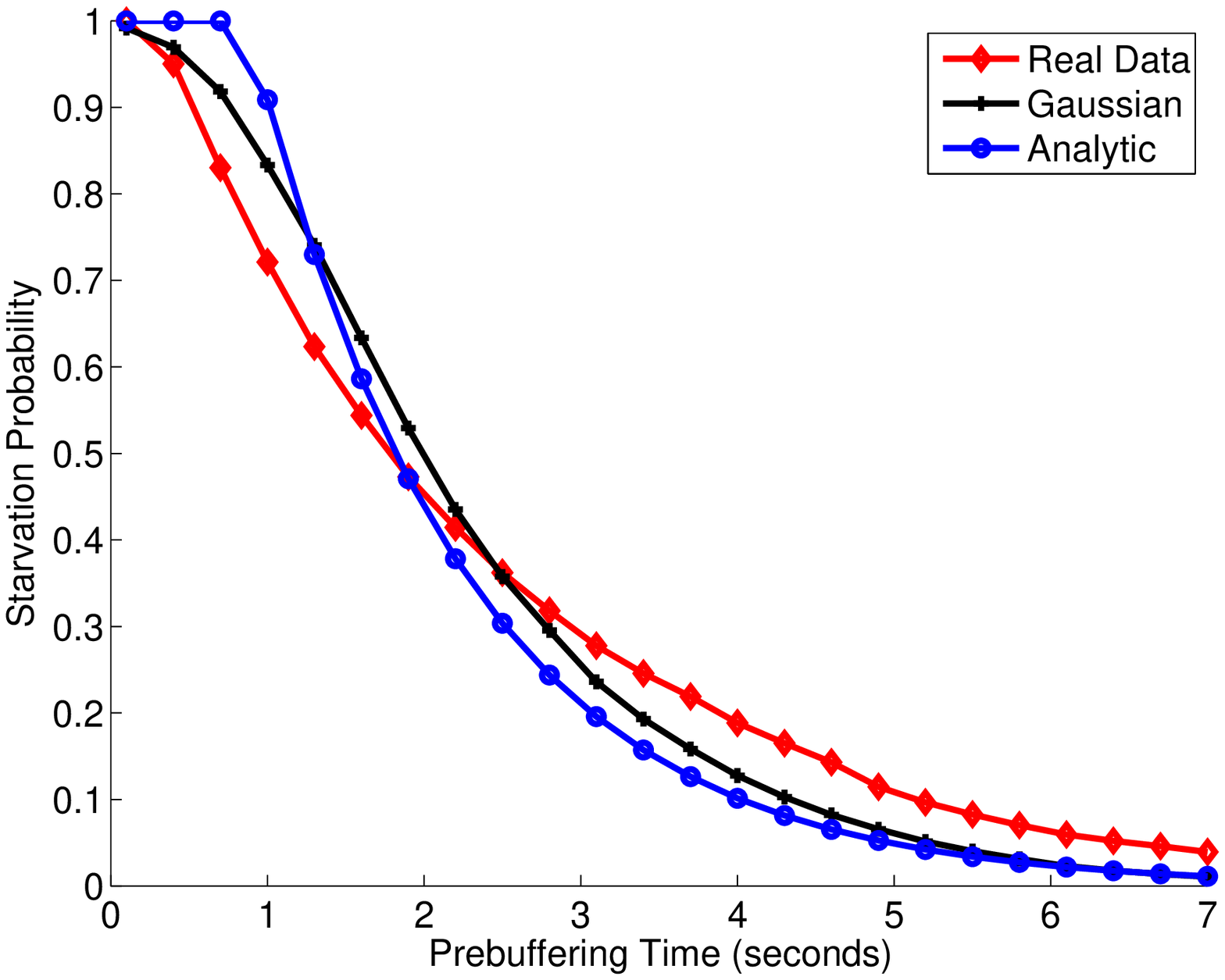}}
    \subfigure{\includegraphics[width=\figsizeA]{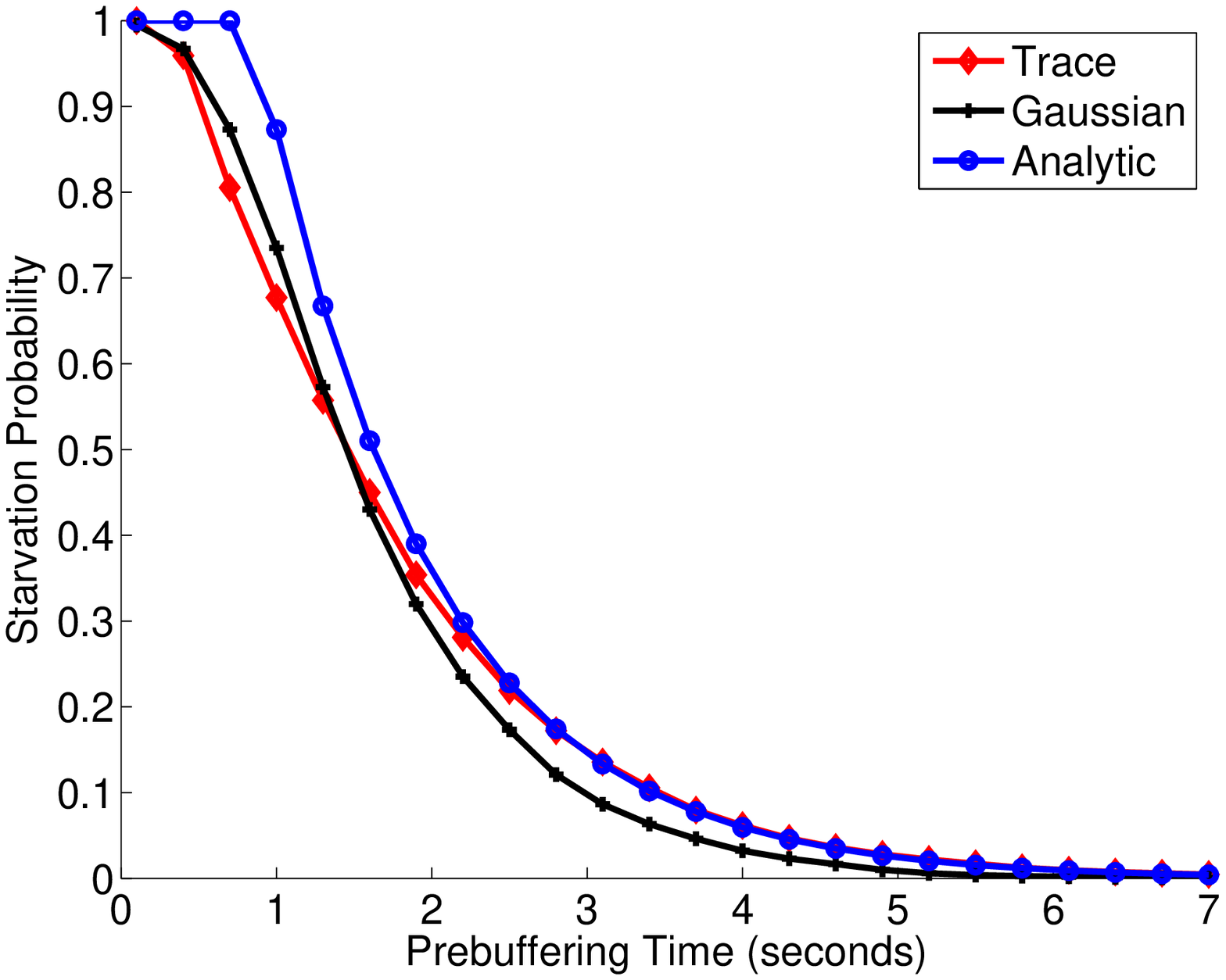}}
	\subfigure{\includegraphics[width=\figsizeA]{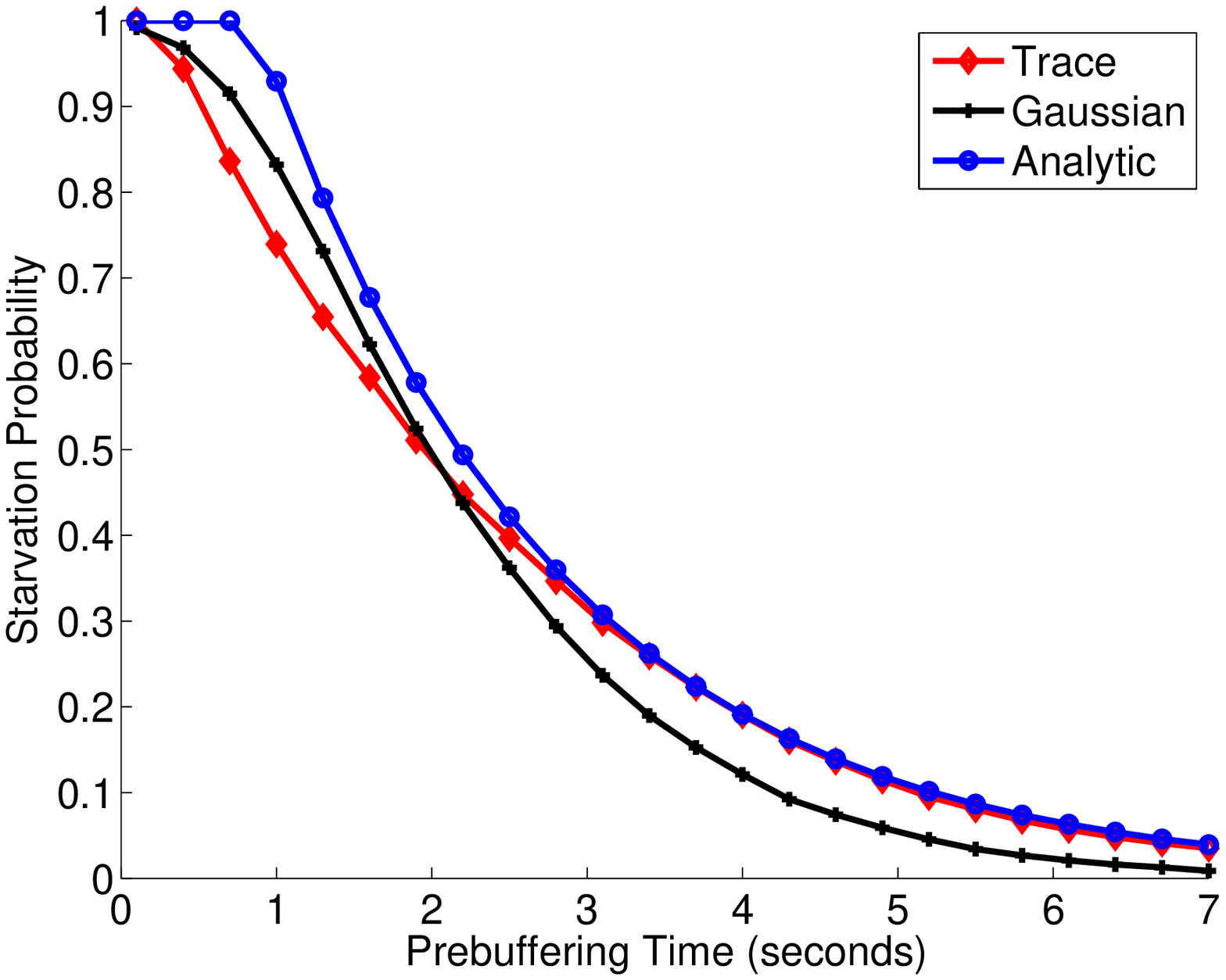}}
    \subfigure{\includegraphics[width=\figsizeA]{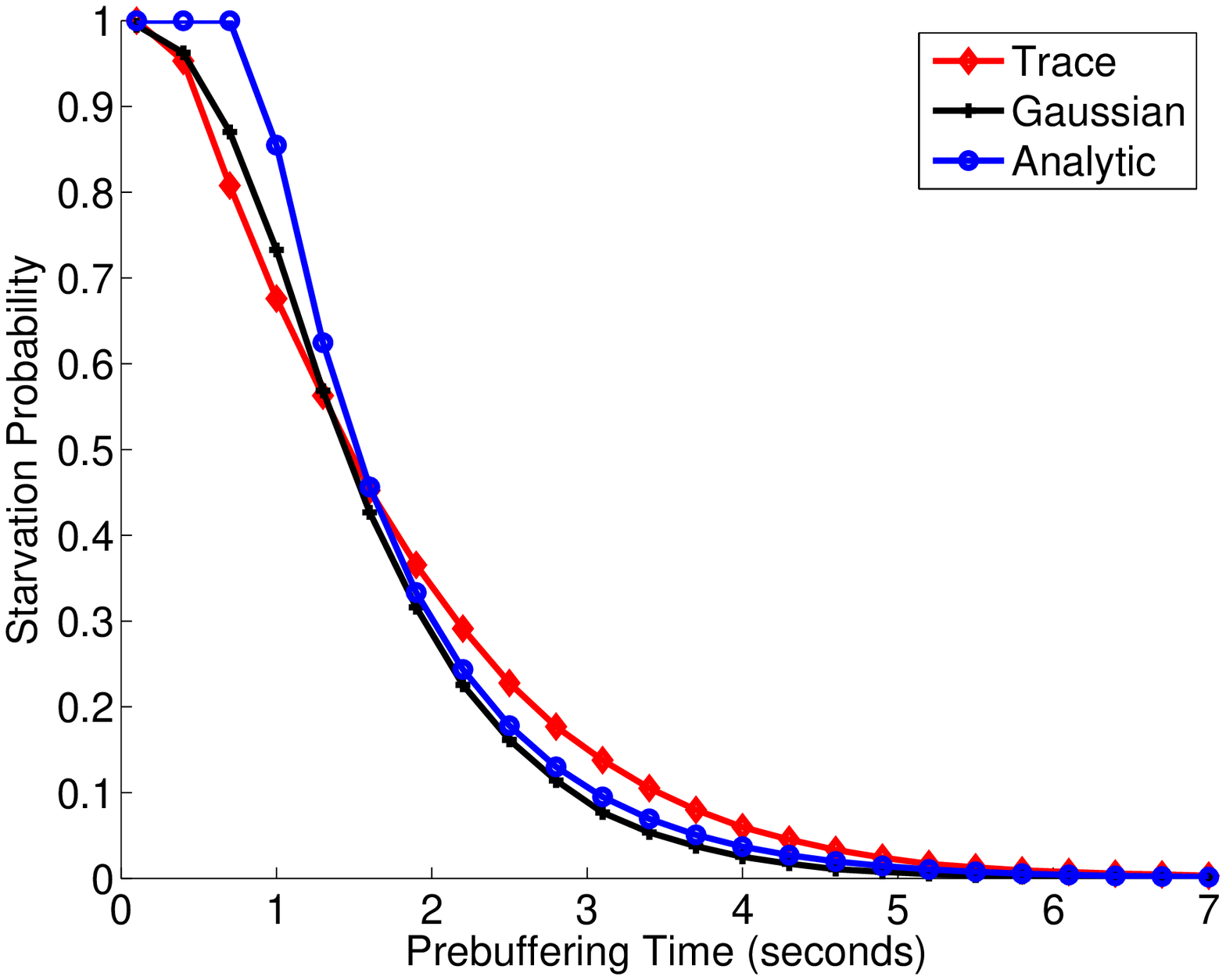}}
\end{center}
\caption{Real-world data: starvation probability for different traces.} \label{fig8}
\end{figure}

In Figure~\ref{fig8} we present the starvation probability calculated on four arbitrairly chosen traces. Given a trace, for each link we calculate the empirical distribution of delays, then draw $N$ samples with replacement from this distribution and check whether or not starvation has occured. We then estimate the starvation probability by averaging the result over $10^4$ independent trials (curve 'Trace'). Then we compute $(\hat \mu_k,\hat \sigma_k^2)$ the empirical mean and variance of the delays and we compute the starvation probability when delays are i.i.d. Gaussian with mean and variance $(\hat \mu_k,\hat \sigma_k^2)$ (curve 'Gaussian'). Finally we calculate the analytic formula obtained in Theorem~\ref{th:iid_upper}, statement (ii), when delays are i.i.d. Gaussian with mean and variance $(\hat \mu_k,\hat \sigma_k^2)$ (curve 'Analytic'). The three curves are close to each other so that: i.i.d. Gaussian delays provide a simple and robust model, our analytical formulas predict the starvation probability accurately and provide simple, efficient rules to set the prebuffering time. Delays cannot strictly be Gaussian since they are always positive, but the obtained predictions are accurate across all the considered traces.

\begin{table}[!htbp]
\begin{center}
\begin{tabular}{|c|c|c|c|c|c|c|c|c|}
\hline
Lag & 0 & 1 & 2 & 3 & 4 & 5 & 6 & 7  \\
 \hline 
Link 1   &  1.00   &  0.04   &  0.04   &  0.06   &  0.06   &  0.06   &  0.05   &  0.05    \\ 
 \hline 
Link 2   &  1.00   &  0.03   &  0.04   &  0.04   &  0.04   &  0.05   &  0.05   &  0.06   \\ 
\hline
\end{tabular}
\end{center}
\caption{Real-world data: autocorrelation of delays}\label{tab1}
\end{table}
\section{Conclusion}\label{sec:conclusion}
We have investigated streaming over multiple links. We have provided lower bounds on the starvation probability of any policy and proposed simple schemes that approach those limits. For general delay distributions, we have provided tractable upper bounds for the starvation probability of the proposed policies. Our results cover several models of practical interest such as links that employ CSMA or opportunistic scheduling at the packet level, on-off channels and links that use fair rate sharing at the flow level. Numerical experiments demonstrate the accuracy of the proposed bounds and approximations.

\section*{Acknowlegements}
This work has been performed in the framework of the IDEFIX project, funded by the ANR (Agence Nationale de la Recherche) under the contract number ANR-13-INFR-0006.

%\newpage
\bibliographystyle{IEEEtran}
\bibliography{main_bibfile}

\begin{table}[!htbp]
\begin{center}
\begin{tabular}{|l l|}
\hline
   & \hspace{-1.4cm} \underline{General} \\
   $N$ & number of chunks per file \\
   $K$ & number of links\\
   $B$ & pre-buffering time\\
   $d_k(n)$ & number of chunks $\le n$ requested on link $k$ \\
   $X_k(\ell)$ & delay of the $\ell$-th chunk requested on link $k$ \\
   $P$ & starvation probability \\
   $\pi$ & chunk request policy \\
   $\mu_k$ & expected delay of link $k$ \\
   $r_k$ & expected data rate of link $k$ \\
   $R$ & sum rate of links \\
   $f_k$ & frequency of link $k$ \\
   $\Psi$ & c.c.d.f. of the standard normal distribution \\
    & \hspace{-1.4cm} \underline{I.i.d. delays} \\
   $\sigma_k^2 $ & variance of delays of link $k$\\
   $G_k$ & cumulant generating function of delays of link $k$ \\
   $v_k^2$ & variance upper bound for sub-Gaussian delays \\ 
   $F_k(a)$ & $G_k(a) - a/f_k$ \\
   $a_k^\star$ & largest zero of $F_k$ \\ 
   & \hspace{-1.4cm} \underline{Markovian delays} \\
   $S_k(t) $ & state of link $k$ at time $t$ \\
   $Q^k$ & transition rate matrix of link $k$ \\
   $m_k$ & stationary distribution of link $k$ \\
   ${\cal S} $ & link state space \\
   $r(.) $ & instantaneous data rate \\
   $\phi$ & speed \\
   $g^k$ & solution to the Poisson equation \\
   $\bar{\sigma}^2_k$ & asymptotic variance (from the Poisson equation) \\
& \hspace{-1.4cm} \underline{Wireless links with random access / scheduling} \\
	$T_s$ & time slot duration \\
	$W$ & window size \\
	$p$ & success probability \\
	$n_f$ & number of frames per chunk \\
	& \hspace{-1.4cm} \underline{Wireless ON-OFF channels} \\
	$\alpha,\beta$ & transition probabilities \\
		& \hspace{-1.4cm} \underline{Sharing links with small flows} \\
	$\lambda$ & arrival rate \\
	$\mu$ & service rate \\
	$\rho$ & load \\
	$\bar{r}$ & expected data rate \\
\hline
\end{tabular}
\end{center}
\caption{Used Notation: Index}
\end{table}

\end{document}